\documentstyle[12pt,epsfig]{article}

\newcommand{\lsim}{\raisebox{-0.13cm}{~\shortstack{$<$ \\[-0.07cm] $\sim$}}~}
\newcommand{\gsim}{\raisebox{-0.13cm}{~\shortstack{$>$ \\[-0.07cm] $\sim$}}~}

\newcommand{\ra}{\rightarrow}

\newcommand{\s}{\\ \vspace*{-3.5mm} }
\newcommand{\nn}{\noindent}
 
\newcommand{\non}{\nonumber}
\newcommand{\beq}{\begin{eqnarray}}
\newcommand{\eeq}{\end{eqnarray}}
\newcommand{\tb}{\tan\beta}

\newcommand{\ct}[1]{c_{\theta_#1}}
\newcommand{\st}[1]{s_{\theta_#1}}
\newcommand{\sinb}{\sin\beta}
\newcommand{\cosb}{\cos\beta}

\newcommand{\musq}{\mu_{\tilde q}}

\newcommand{\mui}{\mu_{\chi_i}}

\begin{document}
\def\thefootnote{\fnsymbol{footnote}}

\begin{flushright}
PM/00--41\\
November 2000
\end{flushright}

\vspace{1cm}

\begin{center}

{\large\sc {\bf Three--Body Decays of Top and Bottom Squarks}} 

\vspace{0.9cm}

{\sc A. Djouadi$^1$ and Y. Mambrini$^2$} 

\vspace{0.8cm}

Laboratoire de Physique Math\'ematique et Th\'eorique, UMR5825--CNRS,\\
Universit\'e de Montpellier II, F--34095 Montpellier Cedex 5, France. 

\vspace*{.4cm} 

$^1$ djouadi@lpm.univ-montp2.fr \\
$^2$ yann@lpm.univ-montp2.fr \\

\end{center}

\vspace{2.5cm}

\begin{abstract}
\nn We investigate the decays of third generation scalar quarks in the Minimal
Supersymmetric extension of the Standard Model, focusing on the three--body
modes. We calculate the partial widths of the decays of heavier top and bottom 
squarks into the lighter ones and a fermion pair [through virtual vector boson,
Higgs boson or gaugino exchanges] and the partial widths of the three--body 
decays of both top squarks into bottom quarks and a pair of fermion and scalar 
fermion [we consider the case of lighter $\tilde{\tau}$ or $\tilde{b}$ states]
and  into a bottom quark, the lightest neutralino and a $W$ or a charged Higgs 
boson $H^\pm$. Some of these decay modes are shown to have substantial 
branching ratios in some areas of the parameter space. 
\end{abstract}

\def\thefootnote{\arabic{footnote}}
\setcounter{footnote}{0}

\newpage

\subsection*{1. Introduction} 

In the Minimal Supersymmetric extension of the Standard Model (MSSM) \cite{R1},
the spin--zero partners of third generation standard chiral fermions can be
significantly lighter than the corresponding scalar partners of first and second
generation fermions. This is essentially due to the relatively large values of
third generation fermion Yukawa couplings which enter in the non--diagonal 
entries of the sfermion mass matrices, the diagonalization of which turn the 
left-- and right--handed current eigenstates $\tilde{f}_L$ and $\tilde{f}_R$ 
into the mass eigenstates $\tilde{f}_1$ and $\tilde{f}_2$ \cite{R2}.
The mixing can generate a sizeable splitting between the masses of the two
physical states and leads to a lighter sfermion $\tilde{f}_1$ with a mass
possibly much smaller that the masses of the other sfermions. The situation can
be even more special in the case of the lightest top squark, $\tilde{t}_1$,
whose mass can be smaller than the one of its partner the top quark,
$m_{\tilde{t}_1} \lsim m_t$, to be compared with the experimental lower bound on
the masses of the first and second generation squarks, $m_{\tilde{q}} \gsim 
{\cal O}(250$ GeV) \cite{R3}. \s

The fact that the top quark is heavy leads to distinct phenomenological 
features for the decays of its scalar partners. Indeed, while the other
squarks can decay directly into (almost) massless quarks and the
lightest neutralino $\chi_1^0$, which is always kinematically accessible 
since in the MSSM the neutralino $\chi_1^0$ is assumed to be the lightest 
supersymmetric particle (LSP), the decay channels $\tilde{t}_i \to t \chi_1^0$ 
are kinematically closed for $ m_{\tilde{t}_i} \leq m_t +m_{\chi_1^0}$. 
If, in addition, $ m_{\tilde{t}_i} \leq m_b +m_{\chi_1^+}$ with $\chi_1^+$
being the lightest chargino, the decay modes $\tilde{t}_i \to b \chi_1^+$ 
are not accessible and the only two--body decay channel which would be 
allowed is the loop induced and flavor changing decay into a charm quark and 
the LSP, $\tilde{t}_i \to c \chi_1^0$ \cite{R4}. The other possible mode is 
the four--body decay channel into a bottom quark, the LSP and two massless
fermions, $\tilde{t}_i \to b \chi_1^0 f \bar{f}'$, which occur through
virtual exchange of top quarks, charginos and scalar fermions \cite{R5}. \s

For relatively heavier top squarks, the three--body decay channels 
\beq 
\tilde{t}_i \ \to \ b W^+ \chi_1^0 \ \ , \ \ b H^+ \chi_1^0
\eeq
where $H^\pm$ is the MSSM charged Higgs boson, can be accessible; Fig.~1a-b. 
These decays have been discussed in Ref.~\cite{R6a,R6b} in the case of the 
lightest top squark and have been shown to be [at least for the one with $W$ 
boson final states] often 
dominant in the case where $m_{\tilde{t}_1} \leq m_t +m_{\chi_1^0}$ and $m_b +
m_{\chi_1^+}$. In addition, if sleptons are lighter than squarks [as is often 
the case in models with a common scalar mass at the GUT scale such as the 
minimal Supergravity (mSUGRA) model] the modes\footnote{This mode has
also been discussed in Ref.~\cite{R7} for first and second generation 
slepton decays into lighter $\tau$ sleptons in the context of gauge mediated 
Supersymmetry breaking models \cite{GSMB}.} 
\beq 
\tilde{t}_i \ \to \ b  l^+ \tilde{\nu}_l\ \ {\rm and/or} \ \ b \tilde{l}^+ \nu_l
\eeq
become possible \cite{R4,R6b,R6c}, Fig.~1c. In the case of the lightest top 
squarks, they can be largely dominating over the loop induced $c\chi_1^0$ mode.
\s

In this paper, we point out that these three--body decay modes are important
not only for the lightest top squark, but also for the heavier one. In
addition, we investigate a new possibility which is the decay of the top
squarks into a fermion--antifermion pair and the lightest $\tilde{b}$ state, 
which is mediated by the virtual exchange of $W$ and $H^+$ bosons: 
\beq
\tilde{t}_i \ \to \ \tilde{b}_1 \, f \bar{f}'
\eeq
$\tilde{b}_1$ can become the lightest scalar quark in the case where the ratio
of the vacuum expectation values of the two--Higgs doublet fields in the MSSM,
$\tb$, is large\footnote{The scenario with large values of $\tb$, $\tb \sim
m_t/m_b$, is favored  in models with Yukawa coupling unification at
the GUT scale \cite{R8}; the other possible solution, with $\tb \sim 1.5$,
seems to be ruled out by the negative searches of MSSM Higgs bosons at LEP2
\cite{R9}.}. \s

For the heavier top squark, $\tilde{t}_2$, another possibility would be the
three--body decay into the lightest top squark and a fermion pair [with $f
\neq b$] through the exchange of the $Z$ and the MSSM neutral Higgs bosons 
[the CP--even $h,H$ and the CP--odd $A$ bosons],
\beq
\tilde{t}_2 \ \to \ \tilde{t}_1 \, f \bar{f}
\eeq
These modes apply also for the charged decays of heavier bottom squarks
into top squarks (and vice--versa) which, as in eq.~(3), occur through 
$W$ and $H^+$ boson exchanges\footnote{If the mass splitting between the 
initial and final scalar eigenstates is large enough, the gauge and Higgs 
bosons become real, and we have the two--body decays into gauge and Higgs 
bosons which have been recently analyzed in Ref.~\cite{R10}.}
\beq
\tilde{q}_2 \ \to \ \tilde{q}_j' \, f \bar{f}'
\eeq
For $b\bar{b}$ final states, one needs to include in the case of $\tilde{t}_2 
\to \ \tilde{t}_1 \, b \bar{b}$ the contributions of the exchange of the two
charginos states $\chi_{1,2}^+$; Fig.~1d--e. This is also the case of the decay
mode, $\tilde{b}_2 \ \to \ \tilde{b}_1 \, b \bar{b}$, where one has in
addition, the virtual exchange of neutralinos and gluinos, Fig.~1e, which have
to be taken into account. The latter process is a generalization [since the
mixing pattern is more complicated] of the decay modes of first and second
generation squarks into light scalar bottoms discussed in Ref.~\cite{R11}, and
would be in competition with at least the two--body mode $\tilde{b}_2 \to b
\chi_1^0$. The latter channel is always open since $\chi_1^0$ is the LSP, but
the $b$--$\tilde{b}_2$--$\chi_1^0$ coupling can be small, leaving the
possibility to the three--body mode to occur at a sizeable rate.  \s

In this paper we analyze all the three--body decay modes, eqs.~(2--5),
discussed above. We will give complete analytical expressions for the Dalitz
plot densities in terms of the energies of the final fermions as well as the
fully integrated partial decay widths. In addition, we investigate the
$\tilde{t}_1$ and $\tilde{t}_2$ decay modes of eq.~(1) which have been already
discussed in Ref.~\cite{R6b} for the lightest top squark $\tilde{t}_1$.  In
this case however, only the Dalitz densities will be given; the more complete 
and lengthy formulae can be found in Ref.~\cite{R12}. \s

The rest of the paper is organized as follows. In the next section, we
will discuss the main properties of top and bottom squarks and summarize 
their two--body decay modes for completeness. In sections 3 and 4, we analyze 
respectively, the decay modes of top and bottom squarks into lighter sfermions 
and fermion pairs, and the decays of the two top squarks into neutralinos, $b$ 
quarks and $W$ or $H^+$ bosons. A numerical illustration is given in 
section 5 and a brief conclusion in section 6. 

\begin{figure}[htbp]
\vspace*{-2cm}
\hspace*{-1cm}
\psfig{figure=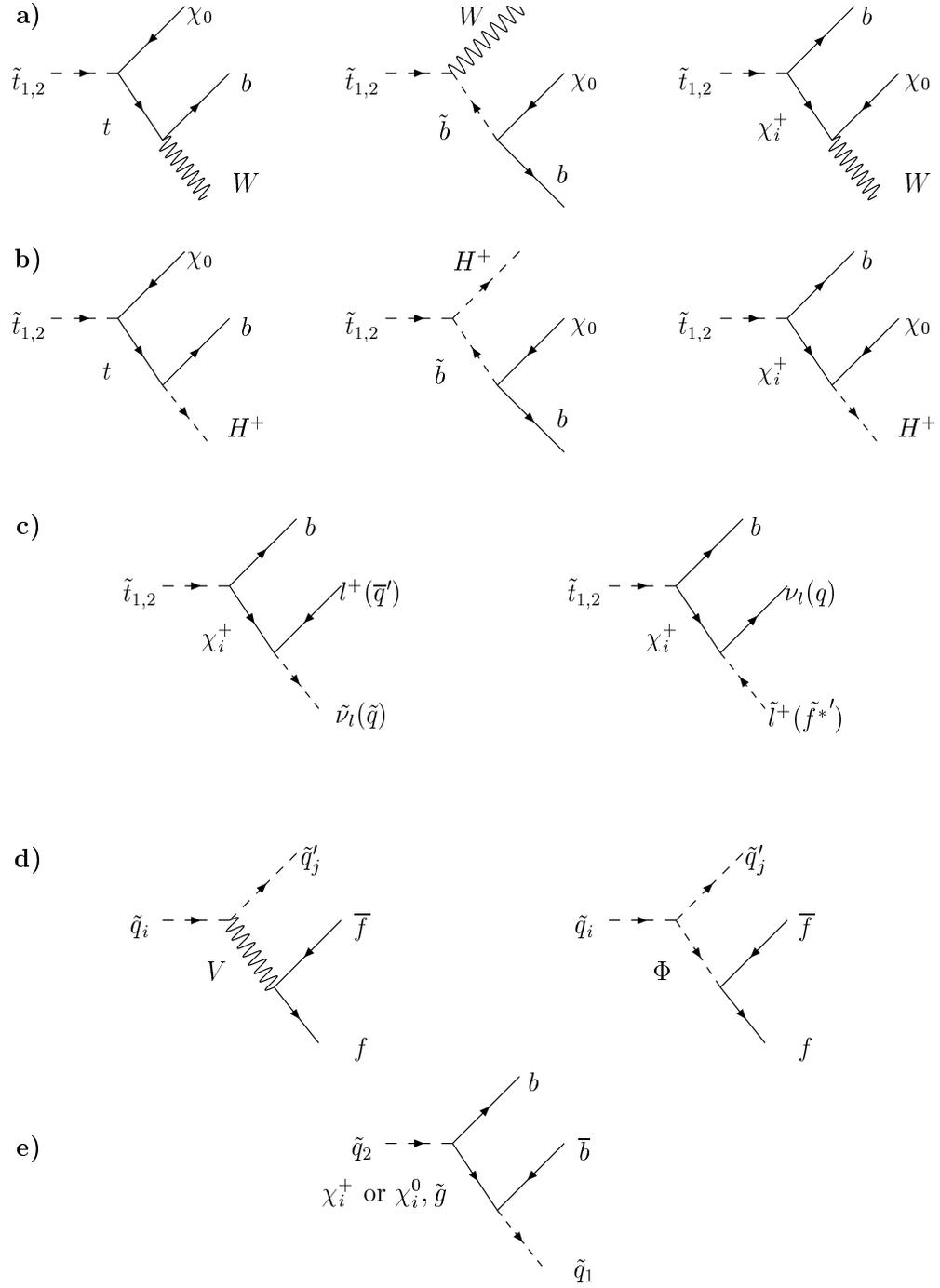,width=19.cm}
\vspace*{-6cm}
\caption[]{Feynman diagrams for the three--body decay modes of top and bottom 
squarks.}
\end{figure}

\subsection*{2. The Two--Body Decay Modes} 

In this section, we will  summarize for completeness the two--body decays of
scalar quarks. This will give us the opportunity to exhibit the various
couplings of squarks to charginos, neutralinos, Higgs and gauge bosons which
will be needed later on, and to discuss the third generation sfermion mass
spectrum and mixing pattern.  

\subsubsection*{2.1 Sfermion masses and mixing}

As mentioned earlier, the left--handed and right--handed sfermions of the
third generation $\tilde{f}_L$ and $\tilde{f}_R$ [the current eigenstates] 
can strongly mix to form the mass eigenstates $\tilde{f}_1$ and $\tilde{f}_2$; 
the mass matrices which determine the mixing are given by
\begin{eqnarray}
 M_{\tilde{f}}^2 \ = \ 
\left[ \begin{array}{cc} m_{LL}^2  & m_f \tilde{A}_f  
\\ m_f \tilde{A}_f & m_{RR}^2  
\end{array} \right]
\end{eqnarray}
with, in terms of the soft SUSY--breaking scalar masses $m_{\tilde{f}_L}$ and  
$m_{\tilde{f}_R}$, the trilinear coupling $A_f$, the higgsino
mass parameter $\mu$ and $\tb =v_U/v_D$, the ratio of the vacuum expectation 
values of the two--Higgs doublet fields 
\begin{eqnarray}
m_{LL}^2&=& m_f^2+m_{\tilde{f}_L}^2 + (I_f^3 -e_f s_W^2)\cos 2\beta\,
M_Z^2  \non \\
m_{RR}^2 &=& m_f^2+m_{\tilde{f}_R}^2 + e_f s_W^2\,\cos 2\beta\,M_Z^2 \non \\
\tilde{A}_f &=& A_f-\mu (\tb)^{-2I_f^3}   \hspace*{3cm} 
\end{eqnarray}
with $e_f$ and $I_f^3$ the electric charge and weak isospin of the 
sfermion $\tilde f$ and  $s_W^2=1-c_W^2\equiv \sin^2\theta_W$. In the
stop sector, the mixing is strong for large values of the trilinear coupling
$A_t$ and/or for large values of $\mu$ with small values of $\tan \beta$.
In the case of the scalar bottom and tau lepton, the mixing is large when
$\tan \beta$ and the parameter $\mu$ are large. \s

The mass matrices eq.~(6) are diagonalized by $ 2 \times 2$ rotation matrices 
of angle $\theta_f$
\beq
\left(  \begin{array}{c} \tilde f_1 \\ \tilde f_2 \end{array} \right)
={\cal{R}}^{\tilde f} 
\left(  \begin{array}{c} \tilde f_L \\ \tilde f_R \end{array} \right) \non 
\eeq
\beq
 {\cal{R}}^{\tilde{f}} &=&  \left( \begin{array}{cc}
     \ct{f} & \st{f} \\ -\st{f} & \ct{f}
  \end{array} \right)  \ \ \ \ , \ \ \ct{f} \equiv \cos \theta_f 
\ \ {\rm and} \ \ \st{f} \equiv \sin \theta_f 
\eeq
The mixing angle $\theta_f$ and the sfermion eigenstate masses are then given 
by 
\begin{eqnarray}
\label{stopmix}
c_{\theta_f} = \frac{- m_f \tilde{A}_f} {\sqrt{ (m_{LL}^2-m_{\tilde {f}_1}^2)^2
+m_f^2 \tilde{A}_f^2} } \ \ , \ \ 
s_{\theta_f} = \frac{m_{LL}^2 - m_{\tilde{f}_1}^2} 
{\sqrt{ (m_{LL}^2-m_{\tilde {f}_1}^2)^2
+m_f^2 \tilde{A}_f^2}} \hspace*{0.8cm}  \\
m_{\tilde{f}_{1,2}}^2 =  \frac{1}{2} \left[ 
m_{LL}^2 + m_{RR}^2 \mp \sqrt{
(m_{LL}^2 - m_{RR}^2)^2 + 4m_f^2 \tilde{A}_f^2 } \right] \hspace*{1cm}
\end{eqnarray}

\subsubsection*{2.2 Two--body decays into neutralinos and charginos}

If the scalar quarks $\tilde{q}_i$ are heavy enough, their main decay modes 
will be into their partner quarks and neutralinos, $\tilde{q}_i \ra q \chi^0_j$ 
[$j$=1--4], and quarks and charginos, $\tilde{q}_i \ra q' \chi^\pm_j$ 
[$j$=1--2]. The partial decay widths are given at the tree--level 
by\footnote{The QCD corrections to these decay modes have been calculated in 
Ref.~\cite{C1}.}:
\begin{eqnarray}
\Gamma( \tilde{q}_i \ra q \chi_j^0) &=& \frac{\alpha \lambda^{\frac{1}{2}} 
(m_{\tilde{q}_i}^2,m_{q}^2,m_{\chi_j^0}^2)}{4m_{\tilde{q}_i}^3}
             \bigg[ ( {a^{\tilde q}_{ij}}^2 + {b^{\tilde q}_{ij}}^2 ) 
                     ( m_{\tilde{q}_i}^2 - m_{q}^2 - m_{\chi_j^0}^2 )
                    - 4 a^{\tilde q}_{ij} b^{\tilde q}_{ij} m_{q} 
             m_{\chi_j^0} \epsilon_{\chi_j} \bigg] \nonumber \\
\Gamma(\tilde{q}_i \ra q' \chi_j^\pm) &=& \frac{\alpha \lambda^{\frac{1}{2}}
(m_{\tilde{q}_i}^2,m_{q'}^2,m_{\chi_j^+}^2)}{4 m_{\tilde{q}_i}^3}
             \bigg[ ( {c^{\tilde q}_{ij}}^2 + {d^{\tilde q}_{ij}}^2 )  
                     ( m_{\tilde{q}_i}^2 - m_{q'}^2 - m_{\chi_j^+}^2 )
                    - 4 c^{\tilde q}_{ij} d^{\tilde q}_{ij} m_{q'} m_{\chi_j^+}
             \bigg] \non \\
\end{eqnarray}
where $\lambda (x,y,z)=x^2+y^2+z^2-2\,(xy+xz+yz)$ is the usual two--body
phase space function and $\epsilon_{\chi_j}$ is the sign of the eigenvalue
of the neutralino $\chi_j^0$. The couplings $a_{ij}$ and $b_{ij}$ for the 
neutral decay are given by 
\begin{eqnarray}
\left\{ \begin{array}{c} a^{\tilde f}_{1j} \\  a^{\tilde f}_{2j} \end{array} 
\right\}
        &=&   -\frac{m_f r_f}{\sqrt{2} M_W s_W}
\left\{ \begin{array}{c} \st{f} \\ \ct{f} \end{array} 
\right\}
         - e^f_{Lj}\, \left\{ \begin{array}{c} \ct{f} \\ -\st{f} \end{array} 
\right\} \nonumber \\
\left\{ \begin{array}{c} b^{\tilde f}_{1j} \\  b^{\tilde f}_{2j} \end{array} 
\right\}
        &=&  -\frac{m_f r_f}{\sqrt{2} M_W s_W }
\left\{ \begin{array}{c} \ct{f} \\ -\st{f} \end{array} 
\right\}
         - e^f_{Rj}\, \left\{ \begin{array}{c} \st{f} \\ \ct{f} \end{array} 
\right\}
\end{eqnarray}
with $r_t= N_{j4}/ \sin \beta$ and $r_\tau=r_b=N_{j3}/\cos \beta$ and
\beq
e^f_{Lj} & = & \sqrt{2}\left[ e_f \; N_{j1}'
              + \left(I_f^3 - e_f \, s_W^2 \right)
                 \frac{1}{c_W s_W}\;N_{j2}' \right]  \nonumber \\ 
e^f_{Rj} & = &-\sqrt{2} \, e_f \left[ N_{j1}'
              - \frac{s_W}{c_W} \; N_{j2}' \right]  
\eeq
while the couplings $c_{ij}$ and $d_{ij}$ for the charged 
decay mode are given, for $\tilde{t}_i$ decays, by: 
\beq
\left\{ \begin{array}{c} c^{\tilde t}_{1j} \\ c^{\tilde t}_{2j} \end{array} \right\} & = &
   \frac{m_b\,U_{j2}}{\sqrt{2}\,M_W s_W \,\cosb}
\left\{ \begin{array}{c} \ct{t} \\ -\st{t} \end{array} \right\}
 \non \\
\left\{ \begin{array}{c} d^{\tilde t}_{1j} \\ d^{\tilde t}_{2j} \end{array} \right\} & = &
   \frac{V_{j1}}{s_W} \,
\left\{ \begin{array}{c} -\ct{t} \\ \st{t} \end{array} \right\}
 + \frac{m_t\,V_{j2}}{\sqrt{2}\,M_W s_W\,\sinb}
\left\{ \begin{array}{c} \st{t} \\ \ct{t} \end{array} \right\}  
\eeq
and 
\beq
\left\{ \begin{array}{c} c^{\tilde b}_{1j} \\ c^{\tilde b}_{2j} \end{array} 
\right\} & = &
   \frac{m_t\,V_{j2}}{\sqrt{2} \,M_W s_W \,\sinb}
\left\{ \begin{array}{c} \ct{b} \\ -\st{b} \end{array} \right\}
 \non \\
\left\{ \begin{array}{c} d^{\tilde b}_{1j} \\ d^{\tilde b}_{2j}
\end{array} \right\} & = &
   \frac{U_{j1}}{s_W} \,
\left\{ \begin{array}{c} -\ct{b} \\ \st{b} \end{array} \right\}
 + \frac{m_b\,U_{j2}}{\sqrt{2}\,M_W s_W \,\cosb}
\left\{ \begin{array}{c} \st{b} \\ \ct{b} \end{array} \right\}  
\eeq
for $\tilde{b}_i$ state decays [in the case of $\tau$ sleptons, one has to 
replace in the previous equations, $b$ by $\tau$ and to set $m_t=0$ and 
$\theta_t=0$]. In these equations, $N$ and $U/V$ are the 
diagonalizing matrices for the neutralino and chargino states \cite{C2} with 
\beq
N'_{j1}= c_W N_{j1} +s_W N_{j2} \ \ \ , \ \ \
N'_{j2}= -s_W N_{j1} +c_W N_{j2} \ . 
\eeq

\smallskip

In the case of top squarks, these decays might be not accessible kinematically, 
and the only allowed two--body decay will be the loop induced and flavor 
changing decay mode into a charm quark and the lightest neutralino, 
$\tilde{t}_i \ra c \chi_j^0$. To a good approximation, the partial decay 
widths [in mSUGRA] are given by \cite{R4}: 
\beq
\Gamma( \tilde{t}_i \ra c \chi_j^0) = \frac{\alpha^3}{64\pi^2} m_{\tilde{t}_i} \, 
\left(1- \frac{m_{\chi_j^0}^2} {m_{\tilde{t}_i}^2 } \right)^2 \,
|e_{Lj}^t|^2\left[\frac{ V^*_{tb} V_{cb} \, m_b^2}{2M_W^2 s_W^2 \cos^2 \beta}
\log \left(\frac{\Lambda_{\rm GUT}^2} {M_W^2} \right) \, \times \, \frac{ 
\Delta_i }{ m_{\tilde{c}_L}^2 - 
m_{\tilde{t}_i}^2 }\right]^2
\eeq
\begin{eqnarray}
&& \Delta_1 =  -c_{\theta_t} (m_{\tilde{c}_L}^2+m_{\tilde{b}_R}^2 + m_{H_1}^2 
+A_b^2)  +s_{\theta_t} m_t A_b \non \\
&& \Delta_2 =  s_{\theta_t} (m_{\tilde{c}_L}^2+m_{\tilde{b}_R}^2 + m_{H_1}^2 
+A_b^2)  + c_{\theta_t} m_t A_b 
\end{eqnarray}
The widths are suppressed by the CKM matrix
element $V_{cb} \sim 0.05$ and the (running) $b$ quark mass squared $m_b^2
\sim (3$ GeV$)^2$, but very strongly enhanced by the term $\log \left(
\Lambda_{\rm GUT}^2/ M_W^2 \right)$ with $\Lambda_{\rm GUT} \simeq 2 \cdot 
10^{16}$ GeV. Assuming proper electroweak symmetry 
breaking, the Higgs scalar mass $m_{H_1}$ can be written in terms of $\mu, 
\tb$ and the pseudoscalar Higgs boson mass $M_A$ as $m_{H_1}^2= M_A^2 \sin^2
\beta - \cos 2\beta M_W^2 -\mu^2$. 

\subsubsection*{2.3 Two--body decays of $\tilde{q}_2$ into gauge and Higgs 
bosons}

If the mass splitting between two squarks of the same generation is large
enough, the heavier squark can decay into a lighter one plus a gauge boson 
$V=W,Z$ or a Higgs boson $\Phi=h,H,A,H^\pm$. The partial decay widths 
are given at the tree--level by\footnote{The QCD 
corrections to these decay modes have also been calculated and can be found
in Ref.~\cite{C3}.}:
\beq
\Gamma(\tilde{q}_i \to \tilde{q}_j' V) &=& \frac{\alpha}{4 m_{\tilde{q}_i}^3
M_V^2} \, g_{\tilde{q}_i \tilde{q}_j' V}^2 \, \lambda^{3/2} (m_{\tilde{q}_i}^2,
 M_V^2, m_{\tilde{q}_j'}^2) \\
\Gamma(\tilde{q}_i \to \tilde{q}_j' \Phi) &=& \frac{\alpha}{4 m_{\tilde{q}_i}^3}
\, g_{\tilde{q}_i \tilde{q}_j' \Phi}^2 \, \lambda^{1/2} (m_{\tilde{q}_i}^2, 
M_\Phi^2, m_{\tilde{q}_j'}^2) 
\eeq
In these equations, the couplings of the Higgs bosons to squarks, 
$g_{\tilde{q}_i \tilde{q}_j' \Phi}$, read in the case of neutral Higgs bosons:
\beq
g_{\tilde{q}_1 \tilde{q}_2 h}= \frac{1}{4s_W M_W} \, \bigg[ 
M_Z^2 s_{2\theta_q} (2I_q^3-4 e_q s_W^2) \sin (\alpha+\beta) 
+2 m_q c_{2\theta_q} (A_q r_2^q + 2I_q^3 \, \mu \, r_1^q) \bigg] 
\eeq
\beq
g_{\tilde{q}_1 \tilde{q}_2 H}= \frac{1}{4s_W M_W} \, \bigg[ 
- M_Z^2 s_{2\theta_q} (2I_q^3-4 e_q s_W^2) \cos (\alpha+\beta) 
+2 m_q c_{2\theta_q} (A_q r_1^q - 2I_q^3 \, \mu \, r_2^q) \bigg] 
\eeq
\beq
g_{\tilde{q}_1 \tilde{q}_2 A}= - g_{\tilde{q}_2 \tilde{q}_1 A}= 
\frac{-m_q}{2s_W M_W} \, \bigg[ \mu +A_q (\tan \beta)^{-2I_q^3} \bigg] 
\eeq
with the coefficients $r^q_{1,2}$ as [$\alpha$ is a mixing angle in the 
CP--even Higgs sector of the MSSM, and at the tree--level, can be expressed 
only in terms of $M_A$ and $\tan \beta$]
\beq
r_{1}^t = \frac{ \sin \alpha}{\sin \beta} \ \  , \ \ 
r_{2}^t = \frac{ \cos \alpha}{\sin \beta} \ \ , \ \
r_{1}^b = \frac{ \cos \alpha}{\cos \beta} \ \ , \ \ 
r_{2}^b = - \frac{ \sin \alpha}{\cos \beta}\;.
\eeq
In the case of the charged Higgs boson, the couplings to squarks are given by
\beq
\label{couplingmatrix}
g_{\tilde{q}_i \tilde{q}_j' H^\pm}= \frac{1}{2s_W M_W} \, \sum_{k,l=1}^2 \  
\left( R^{\tilde{q}} \right)_{ik} \, C_{ \tilde{q}
\tilde{q}' H^\pm }^{kl} \, \left( R^{\tilde{q}'} \right)_{lj}^{\rm T}
\eeq
with the matrix $C_{\tilde{q} \tilde{q}' H^\pm }$ summarizing the couplings 
of the $H^+$ bosons to the squark current eigenstates; it is given by 
\beq
C_{\tilde{t} \tilde{b}H^\pm} &= & \sqrt{2} \, \left( \begin{array}{cc}
m_b^2 \tb + m_t^2/\tb - M_W^2 \, \sin 2\beta & m_b \,(A_b \tb  +\mu) \\
m_t \,(A_t/\tb  +\mu) & 2  \,m_t \,m_b/ \sin2\beta
             \end{array} \right)
\eeq

Turning to the couplings of squarks to the $W$ and $Z$ gauge bosons, one has 
\beq
\label{Z0couplings}
g_{\tilde{q}_1 \tilde{q}_2 Z}
= -g_{\tilde{q}_2 \tilde{q}_1 Z} & = & \frac{2I_q^3 s_{2\theta_q}}{4s_W c_W} 
\eeq
\beq
\label{Wcouplings}
g_{\tilde{q}_i \tilde{q}_j' W} & = & \frac{1}{\sqrt{2} s_W} \left( 
\begin{array}{cc}
     \ct{q} \ct{q'} & - \ct{q} \st{q'} \\ -\st{q} \ct{q'} & \st{q} \st{q'}
  \end{array} \right)  
\eeq
Finally, for the next section, we will need the couplings of the $W,Z$ gauge 
bosons and the four Higgs bosons $h,H,A$ and $H^\pm$ to fermions [$r_{1,2}^f$ 
are defined above]: 
\beq
v_{ffZ}= \frac{2I_f^3- 4e_f s_W^2}{4c_Ws_W} \ , \ 
a_{ffZ}= \frac{2I_f^3}{4c_Ws_W} \ , \ v_{ffW} =a_{ffW} = \frac{1}{2\sqrt{2}s_W} 
\\ 
g_{ffh} =\frac{m_f\, r_{2}^f}{2s_W M_W}   \ , \ 
g_{ffH} =\frac{m_f \, r_{1}^f }{2s_W M_W}   \ , \ 
g_{ffA} =\frac{m_f \, (\tb)^{-2I_f^3}}{2s_W M_W} 
\\
g_{udH^\pm}^S= \frac{m_d \tb + m_u {\rm cot}\beta}{2\sqrt{2} s_W M_W} \ , \
g_{udH^\pm}^P= \frac{m_d \tb - m_u {\rm cot} \beta}{2\sqrt{2} s_W M_W} 
\eeq
and the couplings of $W$ and $H^+$ bosons to chargino/neutralino pairs:
\beq
g^{L}_{\chi^0_i \chi^+_j W^-}=  \frac{1}{\sqrt{2}s_W} [- N_{i4} V^*_{j2}+ 
\sqrt{2} N_{i2} V^*_{j1}] \ \ ,  \ g^{R}_{\chi^0_i \chi^+_j W^-}=  
\frac{1}{\sqrt{2}s_W} [N^*_{i3} U_{j2}+ \sqrt{2} N^*_{i2} U_{j1}]
\eeq
\beq
g^{L}_{\chi^0_i \chi^+_j H^-} &=& \frac{\cos \beta}{s_W} \left[ N^*_{i4} 
V^*_{j1} 
+ \frac{1}{\sqrt{2}} \left( N^*_{i2} + \tan \theta_W N^*_{i1} \right) V^*_{j2} 
\right] \non \\
g^{R}_{\chi^0_i \chi^+_j H^-} &=& \frac{\sin \beta}{s_W} \left[ N_{i3} 
U_{j1} - \frac{1}{\sqrt{2}}\left( N_{i2} + \tan \theta_W N_{i1} \right) U_{j2} 
\right]
\eeq

\newpage

\subsection*{3. Decays into scalar fermion final states} 

In this section, we will analyze the decay modes of top and bottom squarks into
lighter scalar fermions, their quark partners and bottom quarks. We will
neglect for simplicity the masses of the final state fermions [except in the
Yukawa couplings] since, even in the case of the bottom quark and tau lepton,
it is a very good approximation for the $\tilde{t}$ and $\tilde{b}$ masses of
the order ${\cal O} (100$ GeV)  that we are considering. In the case of top
quark final states, the $t$--mass effects have of course to be taken into
account and they can be found in Ref.~\cite{R12}. However, in this case
and for top squarks for instance, the two--body decays $\tilde{t}_i
\to t \chi_1^0$ are kinematically allowed and will dominate over all other
decays.  Thus, throughout this paper, we will not consider $t$--quark final
states and take all the fermions to be massless.  

\subsubsection*{3.1 Scalar top decays into lighter sleptons}

We start by considering the decay of top squarks\footnote{The expressions that
we will write in this section are also valid, with the proper change of the
couplings and masses, in the case of bottom squark decays which occur through
neutralino exchange. This decay is of importance in models where SUSY is broken
by gauge interactions (the so--called GMSB models \cite{GSMB}), and where the
lightest SUSY particle, the gravitino, couples very weakly to matter. The
$\tilde{b}$ states will then mainly decay through this channel (but with
neutralino exchange) into the tau slepton which is in general the
next--to--lightest SUSY particle; see Ref.~\cite{GSMB}.} into lighter sleptons,
$\tilde{t}_i \to b l \tilde{l_j}$ with $\tilde{l}_j$ being either a sneutrino
$\tilde{l}_j \equiv \tilde{\nu}_l$ or a charged slepton eigenstate
$\tilde{l}_{1,2}$ [$l=e,\mu,\tau$], which occurs through the exchange of the
two chargino states, $\chi_{1,2}^\pm$.  In terms of the reduced energies of the
final state particles and the reduced slepton mass defined as:  
\beq
x_1=2 \frac{ p_{\tilde{t}_i} \cdot p_b }{ m_{\tilde{t}_i}^2} \ \ , \ \ 
x_2=2 \frac{ p_{\tilde{t}_i} \cdot p_l }{ m_{\tilde{t}_i}^2} \ \ , \ \ 
x_3=2 \frac{ p_{\tilde{t}_i} \cdot p_{\tilde{l_j}}}{m_{\tilde{t}_i}^2} =2-x_1-x_2
\ \ , \ \ \mu_{\tilde{l}}= \frac{m_{\tilde{l_j}}^2}{m_{\tilde{t}_i}^2}
\eeq
the Dalitz density of the decay mode 
reads: 
\beq
&& \frac{ {\rm d} \Gamma (\tilde{t}_i \to b l \tilde{l_j}) }{ {\rm d}x_1 {\rm d}
x_2} = \frac{\alpha^2} {16 \pi} \, m_{\tilde{t}_i} \sum_{k,l=1}^2 \\
&& \Bigg[ 
(c_{ik}^{\tilde{t}} c_{il}^{\tilde{t}} c_{jk}^{\tilde{l}} c_{jl}^{\tilde{l}}+
d_{ik}^{\tilde{t}} d_{il}^{\tilde{t}} d_{jk}^{\tilde{l}} d_{jl}^{\tilde{l}})
\, {\rm d} G_{1kl}^{\tilde{l}} + \sqrt{\mu_{\chi_k}\mu_{\chi_l}}
(c_{ik}^{\tilde{t}} c_{il}^{\tilde{t}} d_{jk}^{\tilde{l}} d_{jl}^{\tilde{l}}+
d_{ik}^{\tilde{t}} d_{il}^{\tilde{t}} c_{jk}^{\tilde{l}} c_{jl}^{\tilde{l}})
\, {\rm d} G_{2kl}^{\tilde{l}} \Bigg] \non 
\eeq
with the two functions ${\rm d} G_{1kl}^{\tilde{l}}$ and ${\rm d} 
G_{2kl}^{\tilde{l}}$ are given by\footnote{
These functions are the same as the ones appearing in the simpler case of
first and second generation squark decays given in eq.~(5) of Ref.~\cite{R11}.
Note that there is a typographical error in the first part of the latter 
equation: $\mu_{\tilde b}$ has to be replaced by $-\mu_{\tilde b}$; the 
integrated form, eq.~(8) of Ref.~\cite{R11},  based on the correct Dalitz 
density is the same as eq.~(38) of the present paper.} 
\beq
{\rm d}G_{1ij}^{\tilde{f}} &=& \frac{(1-x_1)(1-x_2)-\mu_{\tilde{f}}}
{(1-x_1- \mu_{\chi_i}) (1-x_1- \mu_{\chi_j})} \non \\
{\rm d}G_{2ij}^{\tilde{f}} &=& \frac{x_1+x_2 -1 + \mu_{\tilde{f}}}
{(1-x_1- \mu_{\chi_i}) (1-x_1- \mu_{\chi_j})}
\eeq
Integrating the functions over the phase space, with boundary conditions:
\beq
1-x_1- \mu_{\tilde{f}} \leq x_2 \leq 1 - \frac{ \mu_{\tilde{f}}}{1-x_1} \ \ ,
\ \ 0 \leq x_1 \leq 1- \mu_{\tilde{f}}
\eeq
one obtains the functions $G_{1ij}^{\tilde{f}}$ and $G_{2ij}^{\tilde{f}}$
which read:
\beq
G_{1ij}^{\tilde{f}}&=& \frac{1}{4} \Bigg\{ (\mu_{\tilde{f}}-1)(3\mu_{\tilde{f}}
+3- 2\mu_{\chi_i} -2\mu_{\chi_j}) -2 \frac{\mu_{\tilde{f}}^2}{\mu_{\chi_i} 
\mu_{\chi_j}} \log \mu_{\tilde{f}}  \\
&-& 2 \frac{ (\mu_{\tilde{f}} - \mu_{\chi_i})^2 (\mu_{\chi_i}-1)^2} 
{\mu_{\chi_i} (\mu_{\chi_i}- \mu_{\chi_j})} \log \frac{\mu_{\chi_i}
-\mu_{\tilde{f}}}{\mu_{\chi_i}-1}
-2 \frac{ (\mu_{\tilde{f}} - \mu_{\chi_j})^2 (\mu_{\chi_j}-1)^2} 
{\mu_{\chi_j} (\mu_{\chi_j}- \mu_{\chi_i})} \log \frac{\mu_{\chi_j}
-\mu_{\tilde{f}}}{\mu_{\chi_j}-1} \Bigg\} \non 
\eeq
\beq
G_{2ij}^{\tilde{f}}&=& \frac{1}{\mu_{\chi_i}\mu_{\chi_j}} \Bigg\{ 
\mu_{\tilde{f}} \bigg[ 1+ \mu_{\tilde{f}}-  
\frac{\mu_{\tilde{f}}(\mu_{\chi_i} +\mu_{\chi_j})}{ 2 \mu_{\chi_i} 
\mu_{\chi_j}} \bigg] \log \mu_{\tilde{f}} +\frac{1}{2} (1- \mu_{\tilde{f}})(
\mu_{\tilde{f}} + \mu_{\chi_i}\mu_{\chi_j})  \\
&-& \frac{ \mu_{\chi_j} (\mu_{\tilde{f}} - \mu_{\chi_i})^2 (\mu_{\chi_i}-1)^2} 
{2\mu_{\chi_i} (\mu_{\chi_i}- \mu_{\chi_j})} \log \frac{\mu_{\chi_i}
-\mu_{\tilde{f}}}{\mu_{\chi_i}-1}
- \frac{ \mu_{\chi_i} (\mu_{\tilde{f}} - \mu_{\chi_j})^2 (\mu_{\chi_j}-1)^2} 
{2\mu_{\chi_j} (\mu_{\chi_j}- \mu_{\chi_i})} \log \frac{\mu_{\chi_j}
-\mu_{\tilde{f}}}{\mu_{\chi_j}-1} \Bigg\} \non 
\eeq
In the case where $\chi_j=\chi_i= \chi$ [i.e. for the squared terms], the 
expressions simplify to: 
\beq
G_{1ii}^{\tilde{f}}&=& \frac{1}{4} \Bigg\{ (\mu_{\tilde{f}}-1) \bigg( 5 -6 
\mu_\chi + 5 \mu_{\tilde{f}} - 2\frac{ \mu_{\tilde{f}}} {\mu_\chi} \bigg)
- \frac{2 \mu_{\tilde{f}}^2}{ \mu_\chi^2} \log \mu_{\tilde{f}}  \non \\
&+& 2 \frac{ (\mu_{\tilde{f}} - \mu_{\chi})}{\mu_\chi^2} 
(\mu_{\chi}-1) ( \mu_{\tilde{f}} + \mu_{\tilde{f}} \mu_\chi +\mu_\chi 
-3 \mu_\chi^2) \log \frac{\mu_{\chi}-1}{\mu_{\chi}-\mu_{\tilde{f}}} \Bigg\}\\
G_{2ii}^{\tilde{f}} &=& \frac{1}{\mu_\chi^2} \Bigg\{ \frac{1}{2}
(\mu_{\tilde{f}}-1) (\mu_\chi -2 \mu_\chi^2 - 2\mu_{\tilde{f}}+\mu_{\tilde{f}} 
\mu_\chi ) + \mu_{\tilde{f}} (1+\mu_{\tilde{f}}- 
\frac{\mu_{\tilde{f}}}{\mu_\chi} ) \log \mu_{\tilde{f}}  \non \\
&+& \frac{ (\mu_{\tilde{f}} - \mu_{\chi}) (\mu_{\chi}-1) ( \mu_{\tilde{f}} 
- \mu_\chi^2)}{\mu_\chi}
 \log \frac{\mu_{\chi}-1}{\mu_{\chi}-\mu_{\tilde{f}}} \Bigg\} 
\eeq

\subsubsection*{3.2 Scalar top and bottom decays into lighter squarks
and fermion pairs $f \neq b$} 

The neutral decays $\tilde{q}_2 \to \tilde{q}_1 f\bar{f}$,  with $\tilde{q} = 
\tilde{t}$ or $\tilde{b}$, are mediated only by $Z$ boson and $h,H,A$ boson 
exchanges if the final state fermion is not a partner of the decaying squark. 
The Dalitz density, with the reduced energies $x_1= 2(p_{\tilde{q}_2} \cdot 
p_{f})/ m_{\tilde{q}_2}^2$ and $x_2= 2(p_{\tilde{q}_2} \cdot p_{\bar{f}})/ 
m_{\tilde{q}_2}^2$ and the reduced mass squared $\mu_{\tilde{q}} 
=m_{\tilde{q}_1}^2/m_{\tilde{q}_2}^2$, is given by: 
\beq
\frac{ {\rm d} \Gamma (\tilde{q}_2 \to \tilde{q}_1 f\bar{f})}{ {\rm d}x_1 
{\rm d}x_2}  &=& \frac{\alpha^2 N_C} {8 \pi} \, m_{\tilde{q}_2}  \Bigg[ 
\sum_{\Phi, \Phi'=h,H} g_{\tilde{q}_1 \tilde{q}_2 \Phi} g_{\tilde{q}_1 
\tilde{q}_2 \Phi'} g_{ff \Phi} g_{ff \Phi'} \, {\rm d} F_{\Phi 
\Phi'}^{\tilde{q}_1} \non \\
&&+ g^2_{\tilde{q}_1 \tilde{q}_2 A} g_{ffA}^2 {\rm d}F_{AA}^{
\tilde{q}_1} +4 g_{\tilde{q}_1 \tilde{q}_2 Z}^2 
(v^2_{ffZ}+a^2_{ffZ}) \, {\rm d} F_{ZZ}^{\tilde{q}_1} \Bigg]
\eeq
For the charged decay mode, $\tilde{q}_i \to \tilde{q'}_j f \bar{f'}$ [i.e. 
those of the decays $\tilde{t}_{1,2} \to \tilde{b}_{1,2} f \bar{f}'$ and 
$\tilde{b}_{1,2} \to \tilde{t}_{1,2} f \bar{f}'$ which are allowed by phase 
space], mediated by $W$ and $H^+$ boson [if the final fermion $f$ is not a 
partner of the squark $\tilde{q'}_j$], the Dalitz density reads 
\beq
\frac{ {\rm d} \Gamma (\tilde{q}_i \to \tilde{q'}_j f \bar{f'})}{ {\rm d}x_1 
{\rm d}x_2} & = &\frac{\alpha^2 N_C} {8 \pi} \, m_{\tilde{q}_i}\, 
\Bigg[4 g^2_{\tilde{q}_i \tilde{q'}_jW} (v^2_{ffW}+a^2_{ffW}) \,  {\rm d} 
F_{WW}^{\tilde{q'}_j}  \non \\
&& + g^2_{\tilde{q}_i \tilde{q'}_jH^\pm} \left( (g_{ud H^\pm}^{S})^2 + 
(g_{udH^\pm}^{P})^2 \right) \, {\rm d} F_{H^\pm H^\pm}^{\tilde{q}_1} \Bigg] 
\eeq
where $x_1$ and $x_2$ are as above and the reduced mass is now  
$\mu_{\tilde{q}}= m_{\tilde{q'}_j}^2/m_{\tilde{q}_i}^2$; 
$N_C$ is the color factor of the fermion $f$, $N_C=3$ for quarks and 
$N_C=1$ for leptons. \s 

The two functions for the exchange of gauge bosons and scalar bosons ${\rm d} 
F_{VV}^{\tilde{q}}$ [$V=Z,W$] and ${\rm d}F_{\Phi \Phi}^{\tilde{q}}$ [$\Phi, 
\Phi'=h,H,A, H^\pm$] are given by: 
\beq
{\rm d}F_{VV}^{\tilde{q}}= \frac{(1-x_1)(1-x_2)-\mu_{\tilde{q}}}
{(x_1+x_2-1+ \mu_{\tilde{q}} -\mu_V)^2}
\eeq
\beq
{\rm d}F_{\Phi \Phi'}^{\tilde{q}}= \frac{x_1+x_2-1+\mu_{\tilde{q}} }{
(x_1+x_2-1+ \mu_{\tilde{q}} -\mu_\Phi)(x_1+x_2-1+ \mu_{\tilde{q}} -\mu_{\Phi'})}
\eeq
Integrating over the phase with boundary conditions as in eq.~(37), and using 
the phase space function, 
\beq
\lambda (\mu_X, \mu_Y) = -1+2 \mu_X + 2\mu_Y - (\mu_X-\mu_Y)^2 
\eeq
one obtains the integrated functions\footnote{Note that the function 
$F_{VV}^{\tilde{q}}$ is the same as the one obtained in Ref.~\cite{decays}
for the three--body decays of a heavy Higgs boson into a lighter Higgs boson 
and a fermion-antifermion pair.} for the partial decay widths [which have
to be multiplied by the same factors as in eqs.~(42,43)]  
\beq
F_{VV}^{\tilde{q}} &=& \frac{1}{4} \Bigg\{ \frac{1}{3} (1- \mu_{\tilde{q}})  
\bigg[ 5(1+ \mu_{\tilde{q}}) - 4 \mu_V + \frac{2}{\mu_V} \lambda (\mu_V, 
\mu_{\tilde{q}}) \bigg] + \bigg( \lambda (\mu_V, \mu_{\tilde{q}}) -2 
\mu_{\tilde{q}} \bigg) \log \mu_{\tilde{q}} \non \\
&& +2 (1-\mu_V + \mu_{\tilde{q}} ) \sqrt{\lambda (\mu_V, \mu_{\tilde{q}})}
 {\rm Arctan} \bigg[ \frac 
 { (1- \mu_{\tilde{q}}) \sqrt{\lambda (\mu_V, \mu_{\tilde{q}})}} 
{\mu_V(1- \mu_V+ \mu_{\tilde{q}})- \lambda (\mu_V, \mu_{\tilde{q}}) }
\bigg] \Bigg\}
\eeq
\beq
F_{\Phi \Phi'}^{\tilde{q}} &=&  ( \mu_{\tilde{q}} -1) + 
\frac{1}{2}(1+ \mu_{\tilde{q}} -\mu_{\Phi}-\mu_{\Phi'}) \log \mu_{\tilde{q}}
 \non \\ 
&-& \frac{ \mu_\Phi \sqrt{\lambda (\mu_{\tilde{q}}, \mu_\Phi)}} {\mu_\Phi-
\mu_{\Phi'}} {\rm Arctan} \bigg[ \frac 
 { (1- \mu_{\tilde{q}}) \sqrt{\lambda (\mu_\Phi, \mu_{\tilde{q}})}} 
{\mu_\Phi(1- \mu_\Phi+ \mu_{\tilde{q}})- \lambda (\mu_\Phi, \mu_{\tilde{q}})}
\bigg] \non \\ 
&+& \frac{\mu_{\Phi'} \sqrt{\lambda (\mu_{\tilde{q}}, \mu_{\Phi'})}}{
\mu_{\Phi}-\mu_{\Phi'}} {\rm Arctan}\bigg[ \frac 
 { (1- \mu_{\tilde{q}}) \sqrt{\lambda (\mu_{\Phi'}, \mu_{\tilde{q}})}} 
{\mu_{\Phi'}(1- \mu_{\Phi'}+ \mu_{\tilde{q}})- \lambda (\mu_{\Phi'}, 
\mu_{\tilde{q}})} \bigg] \Bigg\}
\eeq
The latter function reduces in the case where $\Phi=\Phi'$, i.e. for the 
squared terms, to 
\beq
F_{\Phi \Phi}^{\tilde{q}} &=&  2( \mu_{\tilde{q}} -1) + 
\frac{1}{2}(1+ \mu_{\tilde{q}} -2 \mu_\Phi) \log \mu_{\tilde{q}}
 \non \\ 
&-& \frac{ \mu_\Phi (1- \mu_\Phi +\mu_{\tilde{q}})+ \lambda (\mu_{\tilde{q}}, 
\mu_\Phi)}{\sqrt{\lambda (\mu_{\tilde{q}}, \mu_\Phi)}} {\rm Arctan} \bigg[ 
\frac { (1- \mu_{\tilde{q}}) \sqrt{\lambda (\mu_\Phi, \mu_{\tilde{q}})}} 
{\mu_\Phi(1- \mu_\Phi+ \mu_{\tilde{q}})- \lambda (\mu_\Phi, \mu_{\tilde{q}})}
\bigg] 
\eeq

\subsubsection*{3.3 Scalar top and bottom decays into lighter squarks
and $b \bar{b}$ pairs} 

In the case of the decays $\tilde{t}_2 \to \tilde{t}_1 b\bar{b}$, there are 
additional contributions with the exchange of charginos, while in the case of 
the decay  $\tilde{b}_2 \to \tilde{b}_1 b\bar{b}$ one has to include the
contributions of virtual neutralinos and gluinos. The Dalitz density 
eqs.~(42,43) have then to be transformed according to:
\beq
&& \frac{ {\rm d} \Gamma (\tilde{t}_2 \to \tilde{t}_1 b\bar{b})} {{\rm d}x_1 
{\rm d}x_2} \longrightarrow  \frac{ {\rm d} \Gamma (\tilde{q}_2\to \tilde{q}_1 
f\bar{f})} { {\rm d}x_1 {\rm d}x_2} \bigg|_{\tilde{q}=\tilde{t}}^{f=b}
+ \frac{\alpha^2 } 
{16 \pi} \, m_{\tilde{t}_2}\, \Bigg\{ \non \\ 
&& \sum_{k,l=1}^2 \Bigg[
(c_{1k}^{\tilde{t}} c_{1l}^{\tilde{t}} c_{2k}^{\tilde{t}} c_{2l}^{\tilde{t}}+
d_{1k}^{\tilde{t}} d_{1l}^{\tilde{t}} d_{2k}^{\tilde{t}} d_{2l}^{\tilde{t}})
\, {\rm d} G_{1kl}^{\tilde{t}_1} +\sqrt{\mu_{\chi_k^+}\mu_{\chi_l^+}}
(c_{1k}^{\tilde{t}} c_{1l}^{\tilde{t}} d_{2k}^{\tilde{t}} d_{2l}^{\tilde{t}}+
d_{1k}^{\tilde{t}} d_{1l}^{\tilde{t}} c_{2k}^{\tilde{t}} c_{2l}^{\tilde{t}})
\, {\rm d} G_{2kl}^{\tilde{t}_1} \Bigg]  \non \\
&& - 4\sum_{k=1}^2  \Bigg[ g_{\tilde{t}_1 \tilde{t}_2 Z}
\left(
c_{2k}^{\tilde{t}} c_{1k}^{\tilde{t}}(v_{bbZ}-a_{bbZ})+ 
d_{2k}^{\tilde{t}} d_{1k}^{\tilde{t}}(v_{bbZ}+a_{bbZ})
\right)
 \, {\rm d} G_{Vk}^{\tilde{t}_1} \non \\
&&- 2\sum_{\Phi} g_{bb\Phi} g_{\tilde t_1 \tilde t_2 \Phi }
(c_{2k}^{\tilde{t}} d_{1k}^{\tilde{t}}+d_{2k}^{\tilde{t}} 
c_{1k}^{\tilde{t}})\sqrt{\mu_{\chi_k^+} }
\, {\rm d} G_{\Phi k}^{\tilde{t}_1} \Bigg]
\Bigg\} 
\eeq
\beq
&& \frac{ {\rm d} \Gamma (\tilde{b}_2 \to \tilde{b}_1 b\bar{b})} {{\rm d}x_1 
{\rm d}x_2} \longrightarrow  \frac{ {\rm d} \Gamma (\tilde{q}_2\to \tilde{q}_1 
f\bar{f})} { {\rm d}x_1 {\rm d}x_2} \bigg|_{\tilde{q}= \tilde{b}}^{f=b}
+ \frac{\alpha_s^2 } {3 \pi} \, m_{\tilde{b}_2}\, 
\Bigg[ \,2s_{\theta_b}^2 c_{\theta_b}^2
 {\rm d} G_{1\tilde{g} \tilde{g} }^{\tilde{b}_1} 
+ (c_{\theta_b}^4+ s_{\theta_b}^4 ) \mu_{\tilde{g}} 
\,  {\rm d} G_{2\tilde{g} \tilde{g} }^{\tilde{b}_1} \Bigg] \non \\
&& + \frac{\alpha^2 } {16 \pi} \, m_{\tilde{b}_2}\, \Bigg\{ 
\sum_{k,l} \Bigg[ 
(a_{1k}^{\tilde{b}} a_{1l}^{\tilde{b}} a_{2k}^{\tilde{b}} a_{2l}^{\tilde{b}}+
b_{1k}^{\tilde{b}} b_{1l}^{\tilde{b}} b_{2k}^{\tilde{b}} b_{2l}^{\tilde{b}})
\, {\rm d} G_{1kl}^{\tilde{b}_1} +
(a_{1k}^{\tilde{b}} a_{1l}^{\tilde{b}} b_{2k}^{\tilde{b}} b_{2l}^{\tilde{b}}+
b_{1k}^{\tilde{b}} b_{1l}^{\tilde{b}} a_{2k}^{\tilde{b}} a_{2l}^{\tilde{b}})
\non \\
&& \sqrt{ \mu_{\chi_k^0} \mu_{\chi_l^0}}
\, {\rm d}  G_{2kl}^{\tilde{t}_1} \Bigg] -  
4\sum_{k}  \Bigg[ g_{\tilde{b}_1 \tilde{b}_2 Z}
\left[
c_{2k}^{\tilde{b}} c_{1k}^{\tilde{b}}(v_{bbZ}-a_{bbZ})+ 
d_{2k}^{\tilde{b}} d_{1k}^{\tilde{b}}(v_{bbZ}+a_{bbZ})
\right]
 \, {\rm d} G_{Vk}^{\tilde{b}_1} 
\non \\
&&- 2 \sum_{\Phi}g_{bb\Phi} g_{\tilde b_1 \tilde b_2 \Phi}
(c_{2k}^{\tilde{b}} d_{1k}^{\tilde{b}}+d_{2k}^{\tilde{b}} 
c_{1k}^{\tilde{b}})\sqrt{\mu_{\chi_k^0} } \, 
\, {\rm d} G_{\Phi k}^{\tilde{b}_1} \Bigg] \Bigg\} 
\eeq
Note that the sums run only on the virtual states; for instance, in the case
of the decay $\tilde{b}_2 \to \tilde{b}_1 b \bar{b}$ one has to discard the 
exchange of the lightest neutralino $\chi_1^0$ [since it is  the LSP and the 
decay $\tilde{b}_2 \to b \chi_1^0$ always occurs at the two--body level if 
$m_b=0$] and add the two--body partial width $\Gamma(\tilde{b}_1 \to
b \chi_1^0)$ to the total decay width\footnote{If the finite widths of the 
exchanged particles are consistently included in the expressions, one can
use them also for on--shell exchanged particles. However, in the case of the 
decay $\tilde{b}_1 \to b \chi_1^0$ this procedure has always to be done since 
$\chi_1^0$ is stable.}. Note also that the gluino exchange 
diagram does not interfere with the other diagrams due to color conservation.\s

The functions ${\rm d} G_{1kl}^{\tilde{f}} $ and ${\rm d} 
G_{2kl}^{\tilde{f}}$ have been given previously, while 
the new functions ${\rm d} G_{Vi}^{\tilde{f}}$
and ${\rm d} G_{\Phi i}^{\tilde{f}}$ are given by:
\beq
{\rm d}G_{Vi}^{\tilde{q}} &=& \frac{(1-x_1)(1-x_2)-\mu_{\tilde{q}}}
{(x_1+x_2-1+ \mu_{\tilde{q}} -\mu_V)(1-x_1 -\mu_{\chi_i}) } \non \\
{\rm d}G_{\Phi i }^{\tilde{q}}&=& \frac{x_1+x_2-1+\mu_{\tilde{q}}}
{(x_1+x_2-1+ \mu_{\tilde{q}} -\mu_\Phi)(1-x_1 - \mu_{\chi_i})}
\eeq
When integrating over the phase space, one obtains: 
\beq
G_{Vi}^{\tilde{q}} &=&  (\musq-1) \left[ \frac{1}{4} (1+\musq+2\mu_V)
+ (1+\musq-2\mui-2\mu_V)\log\mu_V \right] + \log \frac{\mui-\musq}
{\mui-1} \non \\
&& (\musq-\mui)(\mui-1) 
- \frac{1}{4}\left[ \lambda(\mu_V,\mu_{\tilde q})+2\mui(1-\mu_V+\musq)-
6\musq \right] \log \musq \non \\
&& +\frac{1}{2}\sqrt{\lambda(\mu_V,\mu_{\tilde q})}(1+\musq-2\mui-\mu_V)
{\rm Arctan} \left( \frac{(1-\musq)\sqrt{\lambda(\mu_V,\mu_{\tilde q})}}
{\lambda(\mu_V,\mu_{\tilde q})+\mu_V(\mu_V-1-\musq)} \right) \non \\
&& +\left[ \mui(1+\musq-\mui-\mu_V)-\musq \right] \tilde f(\mu_V, 
r_+^V, r_-^V) 
\eeq
\beq
G_{\Phi i}^{\tilde{q}}= (\musq-1) -\frac{\musq}{ \mu_{\chi_i}} \log {\musq}+ 
\frac{(\musq-\mui)(\mui-1)}{ \mu_{\chi_i}} \log {\frac{\mui-1}{\mui-\musq}} 
+\mu_{\Phi} \tilde f(\mu_\Phi, r_+^{\Phi},r_-^{\Phi})
\eeq
with the function $\tilde{f}$, with arguments $
r_\pm^X= \frac{1}{2} \left[1+\mu_X-\musq \pm \sqrt{-\lambda(\mu_X,\musq)}
\right]$,  defined as:
\beq
\tilde f(z,u,v)= -f(1) + f(u) + f(v)+ \log z \log  
\frac{\mu_{\chi_i} -\mu_{\tilde{q}} }{ \mu_{\chi_i}-1} 
\eeq
where in terms of the Spence functions defined as, ${\rm Li}_2(x)= \int_x^0 
{\rm dt} \, t^{-1} \log (1- t)$, one has:
\beq
f(x)= \log {(\mu_{\chi_i} -1+x)} \log {\frac{1-\mu_{\chi_i}}
{\mu_{\tilde{q}} -\mu_{\chi_i} } } +{\rm Li}_2 \left(\frac{\mu_{\chi_i} -
\mu_{\tilde{q}} } { \mu_{\chi_i} -1+x} \right) -{\rm Li}_2\left(  
\frac{\mu_{\chi_i} -1}{ \mu_{\chi_i} -1+x}\right) 
\eeq

\subsection*{4. Scalar top decays into $W,H^+$ bosons and the LSP}

In this section we will analyze the three--body decay modes of top squarks,
$\tilde{t}_i \to b \chi_j^0 B$ with $B=W,H^+$.  This is a (straightforward)
generalization of the modes $\tilde{t}_1 \to b \chi_1^0 W, H^+$ discussed in
Refs.~\cite{R6a,R6b} since here, we will consider both top squarks in the
initial state and any neutralino in the final state. Here again, we will
neglect the $b$--quark mass in the amplitude squared and in the phase space, as
well as the finite widths of the exchanged particles [the latter can be easily
included in the propagators]; the complete expressions with a finite $m_b$
value in the propagators and in the phase space [which gives a better
approximation for stop masses of order 100 GeV] can be found in
Ref.~\cite{R12}. \s

In terms of the reduced energies of the final particles $x_1=2
(p_{\tilde{t}_i}\cdot p_b)/ m_{\tilde{t}_i}^2$, $x_2= (p_{\tilde{t}_i} \cdot
p_{\chi_j})/m_{\tilde{t}_i}^2$ and $x_3= (p_{\tilde{t} _i} \cdot
p_B)/m_{\tilde{t}_i}^2$, and for the reduced masses $\mu_{X}= p_{X}^2/
m_{\tilde{q}_i}^2$ [we will drop the index for $\chi^0_j$, $\mu_{\chi} \equiv
\mu_{\chi_j^0}$], and introducing the new scaled variables: 
\beq
y_1= \frac{p_b \cdot p_{\chi_j}} {m_{\tilde{q}_i}^2} \  \ , \ \  
y_2= \frac{p_b \cdot p_{B}}{ m_{\tilde{q}_i}^2} \ \  , \ \ 
y_3= \frac{p_{\chi_j} \cdot p_{B}}{m_{\tilde{q}_i}^2} 
\eeq
the Dalitz densities for the decay modes $\tilde{t}_i \to b \chi_j^0 B$ are 
given by:
\beq
\frac{{\rm d}\Gamma}{{\rm d}x_1 {\rm d}x_2} = \frac{\alpha^2}{16 \pi} \, 
\left[ \Gamma^B_{\tilde{b}\tilde{b} } + \Gamma^B_{\chi \chi} +\Gamma^B_{tt} 
+ 2 \Gamma^B_{\tilde{b}\chi}+ 2 \Gamma^B_{\tilde{b} t } + 2\Gamma^B_{\chi t} 
\right]
\eeq
where the terms correspond to the square of the contributions of the sbottom, 
chargino and top quark exchange diagrams and the interference terms. 

\subsubsection*{4.1 The decay $\tilde{t}_i \to   b \chi_j^0 W$}

In the case of the decay $\tilde{t}_i \to b \chi_j^0 W$, one has for the 
various terms: 
\beq
\Gamma^W_{\tilde{b} \tilde{b} }= 8 \sum_{k,l=1}^{2} C_{kl}^0 \frac{ y_1 [ 
\mu_W^{-1} (y_2+y_3)^2 -\mu_{\chi} - 2 y_1]} {(1-x_3 +\mu_W - \mu_{\tilde{b}_k})
(1-x_3 +\mu_W - \mu_{\tilde{b}_l})}
\eeq
\beq
\Gamma^W_{\chi \chi } &=&  \sum_{k,l=1}^{2} \frac{2}{(1-x_1 - 
\mu_{\chi^+_k}) (1-x_1 - \mu_{\chi^+_l})} \Bigg\{ \non \\
&& C_{kl}^{1+}  \bigg[ 4y_3 (y_1+y_2  + \mu_W^{-1} y_1 y_3) + y_1 (\mu_{\chi} 
-\mu_W)  +  2 \mu_{\chi}y_2 (1- \mu_{W}^{-1} y_3) 
 \bigg] \non \\ 
&+&  C_{kl}^{1-} \sqrt{\mu_{\chi^+_k} \mu_{\chi^+_l} }
(y_1+ 2 \mu_W^{-1} y_2 y_3) -3 \sqrt{\mu_{\chi}} (y_1+y_2) 
\bigg[ \sqrt{\mu_{\chi^+_l}} C_{kl}^{2} + \sqrt{\mu_{\chi^+_k}} C^2_{lk} 
\bigg] \Bigg\} 
\eeq
\beq
\Gamma^W_{tt} &=&  \frac{2}{(1-x_2 +\mu_{\chi} - \mu_{t})^2} \Bigg\{ 
C^{3+}_{ij}  \bigg[ 4 y_2 y_1 (\mu_W^{-1} y_2 +1) -\mu_W y_1
+4 y_2 y_3 \bigg] 
\non \\
&+& C^{3-}_{ij} \mu_t [y_1 +2 y_2 y_3 \mu_W^{-1}] 
- 4 \sqrt{\mu_t \mu_\chi} C^4_{ij} y_2 [3+ 2 \mu_{W}^{-1} y_2 ] \Bigg\}
\eeq
\beq
\Gamma^W_{\tilde{b} \chi}&=& 
\sum_{k,l=1}^2 \frac{-4}{(1-x_3+\mu_W-\mu_{\tilde{b}_k})
(1-x_1-\mu_{\chi^+_l})} \Bigg\{ C_{k}^5 \Bigg[(y_2+y_3)(\mu_\chi y_2- 2 y_1 
y_3) \mu_W^{-1} \non \\
&+& y_1(2y_1+y_2-y_3+\mu_\chi) +\mu_\chi y_2 \Bigg] + C_{k}^6 \sqrt{\mu_\chi 
\mu_{\chi^+_l} } \bigg[ y_1 - \mu_W^{-1} y_2(y_2+y_3) \bigg] \Bigg\} 
\eeq
\beq
\Gamma^W_{\tilde{b} t}&=& 
\sum_{k=1}^2 \frac{4}{(1-x_3+\mu_{\chi} -\mu_{\tilde{b}_k})
(1-x_2+\mu_{\chi}-\mu_t)} \Bigg\{ 
\sqrt{\mu_t \mu_\chi} C^{7-}_k \, \Bigg[ y_1 -y_2\mu_W^{-1} (y_2+y_3) \Bigg] 
\non \\
&+& C_{k}^{7+} \Bigg[ y_1 y_2 \bigg( -1+ 2 (y_2+y_3) \mu_W^{-1} \bigg) +
\mu_\chi y_2 - y_1 y_3 -2 y_1^2 \Bigg] \Bigg\} 
\eeq
\beq
\Gamma^W_{\chi t}&=& 
\sum_{k=1}^2 \frac{-2}{(1-x_1 -\mu_{\chi_k^+})(1-x_2+\mu_{\chi}-\mu_t)}
\Bigg\{  \sqrt{ \mu_t \mu_{\chi_k^+} } C_{k}^{8-} (y_1+2 \mu_W^{-1} y_2 y_3)
\non \\
&+ &C^{8+}_k \Bigg[ y_1(2y_3+2y_2+4y_1- \mu_W)+y_2(4y_3+\mu_\chi)-2
\mu_W^{-1} y_2(2y_1 y_3 -\mu_\chi y_2) \Bigg] \non \\ 
&-& 3 \sqrt{ \mu_t \mu_{\chi} } C_{k}^{9+} (y_1+y_2) 
- \sqrt{ \mu_\chi \mu_{\chi^+_j} } C_{k}^{9-} y_2 (3 + 2 \mu_W^{-1} y_2) 
\Bigg\}
\eeq
The various combinations of couplings $C_{0..9}$ read as follows:
\beq
C_{lk}^{0}&=&g_{ \tilde{t}_i \tilde{b}_k W }g_{ \tilde{t}_i 
\tilde{b}_lW} 
(a^{\tilde b}_{kj}a^{\tilde b}_{lj}+b^{\tilde b}_{kj}b^{\tilde b}_{lj})
\non \\  C_{lk}^{1+}&=&d^{\tilde t}_{ik}d^{\tilde t}_{il}G^L_{jkW}G^L_{jlW}
+c^{\tilde t}_{ik}c^{\tilde t}_{il}G^R_{jkW}G^R_{jlW}
\non \\  C_{lk}^{1-}&=&d^{\tilde t}_{ik}d^{\tilde t}_{il}G^R_{jkW}G^R_{jlW}
+c^{\tilde t}_{ik}c^{\tilde t}_{il}G^L_{jkW}G^L_{jlW}
\non \\  C_{lk}^{2}&=&d^{\tilde t}_{ik}d^{\tilde t}_{il}G^L_{jlW}G^R_{jkW}
+c^{\tilde t}_{ik}c^{\tilde t}_{il}G^R_{jlW}G^L_{jkW}
\non \\  C_{lk}^{3 \pm}&=&(a^{\tilde t}_{ij})^2(a_{ffW} \pm v_{ffW})^2+ 
(b^{\tilde t}_{ij})^2(a_{ffW} \mp v_{ffW})^2
\non \\  C_{lk}^{4}&=&a^{\tilde t}_{ij}b^{\tilde t}_{ij}(a_{ffW}^2+v_{ffW}^2)
\non \\  C_{k}^{5}&=&(a^{\tilde b}_{kj}d^{\tilde t}_{il}G^L_{jlW}
+b^{\tilde b}_{kj}c^{\tilde t}_{il}G^R_{jlW})g_{ \tilde{t}_i \tilde{b}_k W }
\non \\  C_{k}^{6}&=&(b^{\tilde b}_{kj}c^{\tilde t}_{il}G^L_{jlW}
+a^{\tilde b}_{kj}d^{\tilde t}_{il}G^R_{jlW})g_{ \tilde{t}_i \tilde{b}_k W }
\non \\  C_{k}^{7+}&=&\bigg[ a^{\tilde b}_{kj}a^{\tilde t}_{ij}(v_{ffW}+a_{ffW})
+b^{\tilde b}_{kj}b^{\tilde t}_{ij}(v_{ffW}-a_{ffW}) \bigg]
g_{ \tilde{t}_i \tilde{b}_k W }
\non \\  C_{k}^{7-}&=&\bigg[ b^{\tilde b}_{kj}a^{\tilde t}_{ij}(v_{ffW}-a_{ffW})
+a^{\tilde b}_{kj}b^{\tilde t}_{ij}(v_{ffW}+a_{ffW})\bigg]
g_{ \tilde{t}_i \tilde{b}_k W }
\non \\  C_{k}^{8 +}&=&a^{\tilde t}_{ij}d^{\tilde t}_{ik}G^L_{jkW}
(v_{ffW}+a_{ffW})
+b^{\tilde t}_{ij}c^{\tilde t}_{ik}G^R_{jkW}(v_{ffW}-a_{ffW})
\non \\  C_{k}^{8-}&=&a^{\tilde t}_{ij}c^{\tilde t}_{ik}G^L_{jkW}
(v_{ffW}-a_{ffW})
+b^{\tilde t}_{ij}d^{\tilde t}_{ik}G^R_{jkW}(v_{ffW}+a_{ffW})
\non \\  C_{k}^{9+}&=&a^{\tilde t}_{ij}c^{\tilde t}_{ik}G^R_{jkW}
(v_{ffW}-a_{ffW})+
b^{\tilde t}_{ij}d^{\tilde t}_{ik}G^L_{jkW}(v_{ffW}+a_{ffW})
\non \\  C_{k}^{9-}&=&a^{\tilde t}_{ij}d^{\tilde t}_{ik}G^R_{jkW}
(v_{ffW}+a_{ffW})+
b^{\tilde t}_{ij}c^{\tilde t}_{ik}G^L_{jkW}(v_{ffW}-a_{ffW})
\eeq 

Note that there are two typographical errors\footnote{We thank Werner Porod for
his cooperation in resolving this issue.} in the corresponding expressions
for these amplitudes in terms of the four--momenta, given in Ref.~\cite{R6a}. 
In eq.~(18) for the amplitude square of the chargino exchange contribution,
the term $2 m_{\chi_1^0}^2$ should be absent in the first square bracket. 
Furthermore, in the first square bracket of eq.~(19) for the amplitude 
squared of the top quark exchange diagram, $4/M_W^2 \times (p_{\chi_1^0} \cdot 
p_W)(p_b \cdot p_W)^2$ should be replaced by $4/M_W^2 \times (p_{\chi_1^0} 
\cdot p_b)(p_b \cdot p_W)^2$, $+3M_W^2 (p_{\chi_1^0} \cdot p_b)$ by $-M_W^2 
(p_{\chi_1^0} \cdot p_b)$ and $-4 (p_b \cdot p_W) (p_{\chi_1^0} \cdot p_W)$
should be replaced by $+4 (p_b \cdot p_W) (p_{\chi_1^0} \cdot p_W)$. [For the
numerical analysis, the agreement with the figures given in Ref.~\cite{R6a} 
is rather good.] 

\subsubsection*{4.2 The decay $\tilde{t}_i \to   b \chi_j^0 H^+$}

In the case of the decay $\tilde{t}_i \to b \chi_j^0 H^+$, one has for the 
various terms: 
\beq
\Gamma^{H^+}_{\tilde{b} \tilde{b} }= 2 \sum_{k,l=1}^{2} D_{kl}^0 \frac{ y_1 }
 {(1-x_3 +\mu_H - \mu_{\tilde{b}_k})
(1-x_3 +\mu_H - \mu_{\tilde{b}_l})}
\eeq
\beq
\Gamma^{H^+}_{\chi \chi } &=&  \sum_{k,l=1}^{2} \frac{2}{(1-x_1 - 
\mu_{\chi^+_k}) (1-x_1 - \mu_{\chi^+_l})} \Bigg\{ 
 D_{kl}^{1-} \sqrt{\mu_{\chi^+_k} \mu_{\chi^+_l} } y_1 \\
&& + D_{kl}^{1+}  \bigg[ 2y_3y_2 + y_1 (\mu_{\chi} 
-\mu_H)  +  2 \mu_{\chi}y_2  \bigg] + \sqrt{\mu_{\chi}} (y_1+y_2) 
\bigg[ \sqrt{\mu_{\chi^+_l}} D_{kl}^{2} + \sqrt{\mu_{\chi^+_k}} D^2_{lk} 
\bigg] \Bigg\}  \non 
\eeq
\beq
\Gamma^{H^+}_{tt}=  \frac{2}{(1-x_2 +\mu_{\chi} - \mu_{t})^2} \Bigg\{ 
D^{3+}_{ij}  (  -\mu_H y_1 +2 y_2 y_3 ) + D^{3-}_{ij} \mu_t y_1 
- 2 \sqrt{\mu_t \mu_\chi} D^4_{ij} y_2  \Bigg\}
\eeq
\beq
\Gamma^{H^+}_{\tilde{b} \chi}&=& -2
\sum_{k,l=1}^2 \frac{ D_{k}^5 \sqrt{\mu_{\chi}} (y_1+y_2) + D_{k}^6 
\sqrt{ \mu_{\chi^+_l} }y_1}   
{(1-x_3+\mu_H-\mu_{\tilde{b}_l}) (1-x_1-\mu_{\chi^+_k})} 
\eeq
\beq
\Gamma^{H^+}_{\tilde{b} t}&=&2 \sum_{k=1}^2 \frac{
\sqrt{\mu_t} D^{7+}_k \,  y_1  - D_{k}^{7-} \,\sqrt{\mu_{\chi}}  y_2  }
{(1-x_3+\mu_{\chi} -\mu_{\tilde{b}_k}) (1-x_2+\mu_{\chi}-\mu_t)} 
\eeq
\beq
\Gamma^H_{\chi t}&=& 
\sum_{k=1}^2 \frac{2}{(1-x_1-\mu_{\chi_k^+})(1-x_2+\mu_{\chi}-\mu_t)}
\Bigg\{  -\sqrt{ \mu_t \mu_{\chi_k^+} } D_{k}^{8-} y_1
\non \\
&+ &D^{8+}_k ( - \mu_H y_1 + 2y_2y_3 +\mu_{\chi}y_2) 
- \sqrt{ \mu_t \mu_{\chi} } D_{k}^{9+} (y_1+y_2) 
+ \sqrt{ \mu_\chi \mu_{\chi^+_k} } D_{k}^{9-} y_2  
\Bigg\}
\eeq
with the various combinations of couplings $D_{0..9}$ given by: 
\beq
 D_{lk}^{0}&=&g_{\tilde{t}_i \tilde{b}_kH}g_{\tilde{t}_i \tilde{b}_lH}
(a^{\tilde b}_{kj}a^{\tilde b}_{lj}+b^{\tilde b}_{kj}b^{\tilde b}_{lj})
\non \\  D_{lk}^{1+}&=&d^{\tilde t}_{ik}d^{\tilde t}_{il}G^L_{jkH}G^L_{jlH}
+c^{\tilde t}_{ik}c^{\tilde t}_{il}G^R_{jkH}G^R_{jlH}
\non \\  D_{lk}^{1-}&=&d^{\tilde t}_{ik}d^{\tilde t}_{il}G^R_{jkH}G^R_{jlH}
+c^{\tilde t}_{ik}c^{\tilde t}_{il}G^L_{jkH}G^L_{jlH}
\non \\  D_{lk}^{2}&=&d^{\tilde t}_{ik}d^{\tilde t}_{il}G^L_{jlH}G^R_{jkH}
+c^{\tilde t}_{ik}c^{\tilde t}_{il}G^R_{jlH}G^L_{jkH}
\non \\  D_{lk}^{3 +}&=&(a^{\tilde t}_{ij})^2(g^S_{tbH^+}+g^P_{tbH^+})^2+ 
(b^{\tilde t}_{ij})^2(g^S_{tbH^+}-g^P_{tbH^+})^2
\non \\  D_{lk}^{3 -}&=&(a^{\tilde t}_{ij})^2(g^S_{tbH^+}-g^P_{tbH^+})^2+ 
(b^{\tilde t}_{ij})^2(g^S_{tbH^+}+g^P_{tbH^+})^2
\non \\  D_{lk}^{4}&=&a^{\tilde t}_{ij}b^{\tilde t}_{ij}
\bigg[ (g^S_{tbH^+}+g^P_{tbH^+})^2+(g^S_{tbH^+}-g^P_{tbH^+})^2 \bigg]
\non \\  D_{k}^{5}&=&(a^{\tilde b}_{kj}d^{\tilde t}_{il}G^L_{jlH}
+b^{\tilde b}_{kj}c^{\tilde t}_{il}G^R_{jlH})g_{\tilde{t}_i \tilde{b}_kH}
\non \\  D_{k}^{6}&=&(b^{\tilde b}_{kj}c^{\tilde t}_{il}G^L_{jlH}
+a^{\tilde b}_{kj}d^{\tilde t}_{il}G^R_{jlH})g_{\tilde{t}_i \tilde{b}_kH}
\non \\  D_{k}^{7+}&=&\left[a^{\tilde b}_{kj}a^{\tilde t}_{ij}
(g^S_{tbH^+}-g^P_{tbH^+})
+b^{\tilde b}_{kj}b^{\tilde t}_{ij}(g^S_{tbH^+}+g^P_{tbH^+})\right]
g_{\tilde{t}_i \tilde{b}_kH}
\non \\  D_{k}^{7-}&=&\left[b^{\tilde b}_{kj}a^{\tilde t}_{ij}
(g^S_{tbH^+}+g^P_{tbH^+})
+a^{\tilde b}_{kj}b^{\tilde t}_{ij}(g^S_{tbH^+}-g^P_{tbH^+})\right]
g_{\tilde{t}_i \tilde{b}_kH}
\non \\  D_{k}^{8 +}&=&a^{\tilde t}_{ij}c^{\tilde t}_{ik}G^R_{jkH}
(g^S_{tbH^+}+g^P_{tbH^+})
+b^{\tilde t}_{ij}d^{\tilde t}_{ik}G^L_{jkH}(g^S_{tbH^+}-g^P_{tbH^+})
\non \\  D_{k}^{8 -}&=&a^{\tilde t}_{ij}d^{\tilde t}_{ik}G^R_{jkH}
(g^S_{tbH^+}-g^P_{tbH^+})
+b^{\tilde t}_{ij}c^{\tilde t}_{ik}G^L_{jkH}(g^S_{tbH^+}+g^P_{tbH^+})
\non \\  D_{k}^{9 +}&=&a^{\tilde t}_{ij}d^{\tilde t}_{ik}G^L_{jkH}
(g^S_{tbH^+}-g^P_{tbH^+})
+b^{\tilde t}_{ij}c^{\tilde t}_{ik}G^R_{jkH}(g^S_{tbH^+}+g^P_{tbH^+})
\non \\  D_{k}^{9 -}&=&a^{\tilde t}_{ij}c^{\tilde t}_{ik}G^L_{jkH}
(g^S_{tbH^+}+g^P_{tbH^+})
+b^{\tilde t}_{ij}d^{\tilde t}_{ik}G^R_{jkH}(g^S_{tbH^+}-g^P_{tbH^+})
\eeq

\newpage

\subsection*{5. Numerical illustrations} 

A Fortran code called {\tt SDECAY} \cite{sdecay} has been developed for the
numerical analysis; all the partial decay widths and the branching ratios for
the two--body and three--body decay modes of scalar quarks [as well as the
decays of charginos, neutralinos, gluinos, sleptons and the four--body decays
of the lightest to squark] have been implemented. It has been interfaced with
the programs {\tt SUSPECT} \cite{suspect} for the calculation of the
supersymmetric particle spectrum [including the renormalization group equations
for the evolution of the SUSY parameters and the implementation of radiative
electroweak symmetry breaking] and the program {\tt HDECAY} \cite{hdecay} for
the Higgs boson spectrum and couplings \cite{C2,coup} where the renormalization
improved two--loop radiative corrections in the MSSM Higgs sector
\cite{two-loop} and the QCD corrections to the Higgs couplings \cite{HQCD} have
been incorporated. \s

We begin our numerical illustration with the decays of the lightest top squark
$\tilde{t}_1$. We will concentrate on the ``unconstrained" MSSM, where for
simplicity, we use a common soft--SUSY breaking scalar mass $m_{\tilde{q}}$ for
the three generations of squarks and $m_{\tilde{l}}$ for the three generations
of sleptons, i.e. $m_{\tilde{t}_L}=m_{\tilde{t}_R}= m_{\tilde{b}_L}
=m_{\tilde{b}_R}= m_{\tilde{q}}$ and $m_{\tilde{\tau}_L}=m_{\tilde{\tau}_R}=
m_{\tilde{\nu}_L}=m_{\tilde{l}}$. [We will also assume that the mixing between
different generations is absent at the tree--level, otherwise the decay mode
$\tilde{t}_i \to c \chi_j^0$ would already occur at this stage.] The mass
splitting between the two mass eigenstates will be then only due to the
different D--terms of $m_{\tilde{f}_L}$ and $m_{\tilde{f}_R}$ and to the
off--diagonal entries of the sfermion mass matrices. The mixing is made strong
in the stop sector by taking large values of $\tilde{A}_t \sim {\cal O}(1\;
{\rm TeV})$; in this case the mixing angle is either close to $\theta_t =\pi/2$
 (no mixing) or to $\pm \pi/4$ (maximal mixing) for respectively small and large
values of the entry $m_t \tilde{A}_t$ compared to the diagonal entries of the
mass matrix. The mixing is strong in the $\tilde{b}$ and $\tilde{\tau}$ sectors
for large values of $\tb$ and the parameter $\mu$, almost independently of
$A_b$ and $A_\tau$ which will be fixed to 100 GeV.   In the gaugino sector, we
will make the usual assumption of the unification of the gaugino masses at the
GUT scale, leading to the relation $M_1 = \frac{5}{3} {\rm tan}^2 \theta_W M_2
\sim \frac{1}{2} M_2$. \s

In Fig.~2, we show the branching ratios of the decays of the lightest top
squark $\tilde{t}_1$ as a function of $\tan \beta$ for large values of
$\mu=-750$ GeV. This implies that the lightest chargino and neutralinos are
gaugino like for $M_2 \lsim 300$ GeV, with masses $m_{\chi_1^+} \simeq
m_{\chi_2^0} \simeq 2 m_{\chi_1^0} \simeq M_2$ [with a very small variation
with $\tb$]. We choose a common squark mass $m_{\tilde{q}}$ of ${\cal O}$(500
GeV) which, for the chosen $\mu$ and $A_t$ values, leads to a $\tilde{t}_1$
with a mass between 170 and 250 GeV [depending on the value of $\tb$,
$m_{\tilde{t}_1} $ being smaller for low $\tb$ values]. In this case,
$m_{\tilde{t}_1}$ is smaller than $m_t+m_{\chi_1^0}$ and $m_b+m_{\chi_1^+}$ but
possibly larger than $m_{\chi_1^0}+M_W, m_{\chi_1^0}+M_{H^+}, m_{\tilde{b}_1}$
or $m_{\tilde{\tau}_1}$, allowing to some three--body decay channels to be open
kinematically. Since these three--body decay modes are of ${\cal O}(\alpha^2)$,
they can compete with the $\tilde{t}_1 \ra c \chi_1^0$ mode which is of ${\cal
O} (\alpha^3)$ modulo the large logarithm $\log(\Lambda_{\rm GUT}^2/M_W^2)$. \s

\setcounter{figure}{1}
\begin{figure}[htbp]
\vspace*{-1.5cm}
\hspace*{-3cm}
\psfig{figure=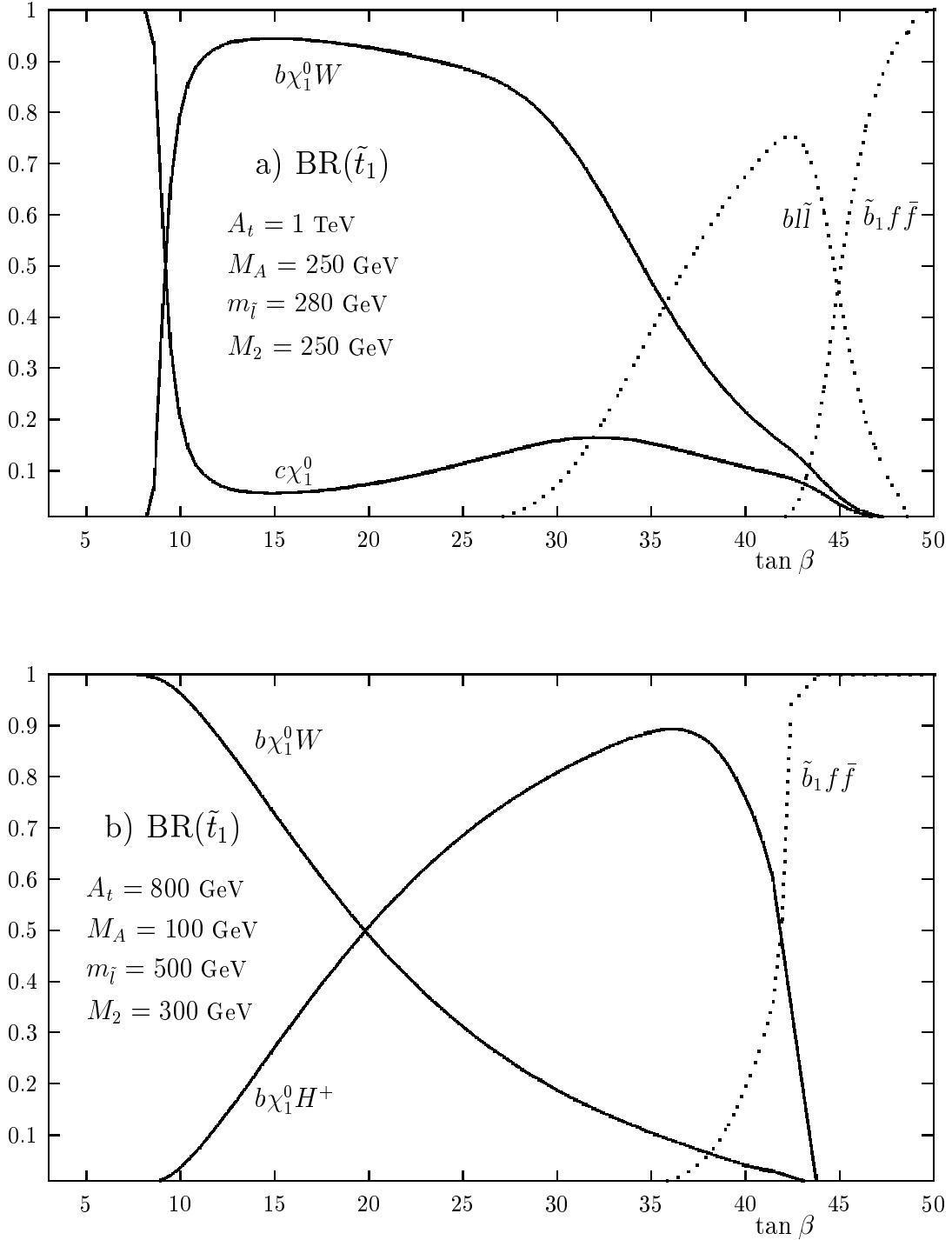,width=21.cm}
\vspace*{-8.cm}
\caption[]{The branching ratios for the two--body and three--body decay modes 
of the lightest top squark $\tilde{t}_1$ as a function of $\tan \beta$ for 
$\mu=-750$ GeV.}
\end{figure}

In Fig.~2a, the common slepton mass is chosen to be relatively small, 
$m_{\tilde{l}}=280$ GeV, to allow for $\tilde{t}_1$ decays into staus and the
pseudoscalar $A$ boson mass is taken to be relatively large, $M_A=250$ GeV,
implying a too heavy charged Higgs boson, $M_{H^+} = \sqrt{M_A^2+M_W^2} 
\simeq 260$ GeV, for the decay $\tilde{t}_1 \to b \chi_1^0 H^+$ to occur. 
$m_{\tilde{q}}$ and $A_t$ are fixed to 450 GeV and 1 TeV, respectively, 
while $M_2=250$ GeV. This leads to a scalar fermion spectrum, for $\tb=5\, 
(45)$, of $m_{\tilde{t}_1} \sim 170\, (230)$ GeV, $m_{\tilde{b}_1} \sim 430\, 
(190)$ GeV and $m_{\tilde{\tau}_1} \sim 270\, (140)$ GeV. \s

For small values of $\tb$, $\tan \beta \lsim 7$, the mixing is too strong in the
stop sector and $\tilde{t}_1$, being too light to have three--body decays, 
will mainly decay into $c\chi_1^0$ final states. For larger values of $\tb$,
the $\tilde{t}_1$ mass becomes larger than $m_b+M_W+m_{\chi_1^0}$ and the 
channel $\tilde{t}_1 \to b \chi_1^0 W$ becomes kinematically accessible; it 
will be largely dominant, with a branching ratio above $\sim 80\%$, up to $\tb 
\sim 30$. For $\tb$ close to the latter value, $\tilde{\tau}_1$ becomes 
relatively light and the decay $\tilde{t}_1 \to b \tilde{\tau}_1^+ \nu$ opens
up and becomes competitive, the branching ratio reaching a maximum at $\tb 
\sim 40$. For even larger values of $\tb$, $\tilde{b}_1$ becomes also light 
and the three--body decay $\tilde{t}_1 \to \tilde{b}_1 f \bar{f}$ will be the 
leading decay channel. For $\tb\gsim 50$, $m_{\tilde{t}_1}$ is larger than 
$M_W+m_{\tilde{b}_1}$ and the two body--decay $\tilde{t}_1 \to \tilde{b}_1 W$ 
opens up and will have a branching ratio close to unity [however, at this 
stage $\tilde{b}_1$ will eventually become lighter than the LSP neutralino]. \s

In the scenario of Fig.~2b, the common slepton mass is taken to be larger than 
previously, $m_{\tilde{l}}=500$ GeV, leading to heavier sleptons [in particular
$\tilde{\tau}$'s] while the pseudoscalar Higgs boson mass is chosen to be 
smaller, $M_A=100$ GeV, leading to a lighter charged Higgs boson, $M_{H^+}
\simeq 126$ GeV, which can thus appear in the decay modes of the $\tilde{t}_1$
state. The other parameters are taken to be $m_{\tilde{q}}=450$ GeV and 
$A_t=800$ GeV, while $M_2$ is fixed to 300 GeV. This gives a scalar fermion 
spectrum, again for $\tb=5\, (45)$, of $m_{\tilde{t}_1} \sim 200 \, (300)$ GeV,
$m_{\tilde{b}_1} \sim 440\, (190)$ GeV and $m_{\tilde{\tau}_1}\sim 500\, (440)$
GeV. \s

In this scenario, since the $\tilde{t}_1$ mass is slightly larger than 
previously, the channel $\tilde{t}_1 \to b\chi_1^0 W$ is already kinematically 
open for small values of $\tb$, and will be the dominating decay mode until 
the channel $\tilde{t}_1 \to b\chi_1^0 H^+$ becomes kinematically accessible. 
The latter will be largely dominating for $20 \lsim \tb \lsim 40$, reaching a 
branching ratio of $\sim 
90\%$ for $\tb \sim 35$, until the opening of the decay channel $\tilde{t}_1 
\to \tilde{b}_1 f\bar{f}$ which becomes accessible for $\tb \sim 35$. This 
channel becomes then quickly the dominant decay mode of $\tilde{t}_1$. Note 
that for $\tb \gsim 45$, $m_{\tilde{t}_1}$ becomes larger than $m_{\tilde{b}_1}
+M_W$ and we have the two--body decay mode $\tilde{t}_1 \to \tilde{b}_1 W$ 
which has a branching ratio very close to unity. \s

Let us now turn our attention to the ``Constrained MSSM'' or minimal 
Supergravity model (mSUGRA) \cite{Manuel} where the soft SUSY breaking scalar 
masses, gaugino masses and trilinear couplings are universal at the GUT scale; 
the left-- and right--handed sfermion masses are then given in terms of the 
gaugino mass parameter $M_{1/2}$, the universal scalar mass  $m_0$, 
the universal trilinear coupling $A_0$ and $\tb$. The soft SUSY breaking 
scalar masses and the trilinear couplings at the low energy scale are given by 
their Renormalization Group Equations, the one--loop approximations of which 
are given for instance in Ref.~\cite{coup,Manuel}. The parameter $\mu$ [up
to its sign] is fixed by the requirement of proper electroweak symmetry 
breaking.
In mSUGRA and in the relatively small $\tan \beta$ regime, due to the running 
of the (large) top Yukawa coupling, the two top squarks can be much lighter 
than the other  squarks, and in contrast with the first two generations one 
has generically a sizeable splitting between $m_{\tilde{t}_L}$ and 
$m_{\tilde{t}_R}$ at the electroweak scale. Thus, even without large mixing, 
$\tilde{t}_1$ can be much lighter than the other squarks in this scenario. \s

In Table 1, we show some of the branching ratios of the lightest top squark for 
the fixed values of the gaugino mass $M_{1/2}=250$ GeV and sign$(\mu)=-$ and 
for several values of the scalar mass $m_0=100, 150, 200$ and $300$ and
several values of $\tb=4,10,20$ and 30 [this leads to $m_{\chi_1^+}
\sim 2  m_{\chi_1^0} \sim 200$ GeV, with a slight dependence on 
$\tb$]. The $\tilde{t}_1$ mass is fixed to approximately $ m_{\tilde{t}_1} 
\sim 200$ GeV by 
varying the trilinear coupling $A_0$. One sees that the branching ratios for 
some of the three--body decays [the channels $\tilde{t}_1 \to b \chi_1^0 
W$ and $b \tilde{\tau}_1 \nu_\tau$] are sizeable. \s

For small values of $\tan \beta$ and the chosen values of $m_0$, 
$\tilde{\tau}_1$ is rather heavy, and the only three-body decay channel which
is available is $\tilde{t}_1 \to b W\chi_1^0$ and for $m_0 =150$ GeV the 
branching ratio is very close to unity. For larger values of $\tb$, 
$\tilde{\tau}_1$ becomes lighter and the phase space for the decay into 
$b W\chi_1^0$ is suppressed so that only the decay mode $\tilde{t}_1 \to b 
\tilde{\tau}_1 \nu_{\tau}$ is largely dominating. For $\tan \beta \gsim 30$,
$\tilde{\tau}_1$ becomes too light [with a mass below the experimental bound]
and the electroweak symmetry breaking does not take place for values of
$m_0, A_t$ leading to a relatively light stop. A few remarks can be made here:
\s

-- For larger values of $M_{1/2}$, the top squarks [and all other squarks] 
become rather heavy and the two--body decays into $b\chi_1^+$ and even to
$t\chi_1^0$ are kinematically allowed and dominate. For smaller values of 
$M_{1/2}$, the chargino  $\chi_1^+$ becomes too light and again, the two--body 
decay channel $\tilde{t}_1 \to b \chi_1^+$ opens up. \s

-- In the studied examples, the parameter $\mu$ is always rather large, $|\mu| 
\gsim 500$ GeV, so that all Higgs particles [except for the $h$ boson] are 
relatively heavy with a mass of ${\cal O}(|\mu|)$. In particular, $H^+$ is too 
heavy for the three--body decay $\tilde{t}_1 \to b H^+ \chi_1^0$ to occur. \s

-- $\tilde{b}_1$ is also rather heavy in the studied scenario.
It is only for very large values of $\tb$ that $\tilde{b}_1$ becomes light
enough for the decay $\tilde{t}_1 \to \tilde{b}_1 f\bar{f}$ to occur. But 
in this case, $\tilde{\tau}_1$ is even lighter and its mass is smaller than the
experimental bound of ${\cal O}(100$ GeV). \s

\begin{table}[hbt]
\renewcommand{\arraystretch}{1.5}
\begin{center}
\begin{tabular}{|c||c||c|c|c|} \hline
\ \ $\tb$ \ \ & \ \ $m_0 $\ \ & BR($b W\chi_1^0)$ & BR$(b \tilde{\sl l}l$) & 
BR$(c \chi_1^0)$ \\  \hline \hline
4    & 150 & 0.993 & $4 \cdot 10^{-2}$  & $3 \cdot 10^{-2}$  \\
10   & 100 & $3 \cdot 10^{-4}$  & 0.915 & 0.085 \\ 
20   & 250 & 0.02 & 0.81 & 0.17 \\ 
30   & 300 & 0.015 & 0.63 & 0.355 \\ \hline
\end{tabular} 
\end{center}
\caption[]{Some examples of branching ratios for the three--body decays of 
$\tilde{t}_1$ in the mSUGRA model for $M_{1/2}=250$ GeV, sign($\mu)=-$ 
and $m_{\tilde{t}_1} \sim 200$ GeV.}
\end{table}

We turn now to the decays of the heavier top squark, $\tilde{t}_2$. In
principle, $\tilde{t}_2$ should have the same decay modes as the lighter 
$\tilde{t}_1$ if the two squarks have approximately the same mass [which means 
that the mixing is not too strong if the left-- and right--handed soft--SUSY
breaking scalar masses, $m_{\tilde{t}_L}$ and  $m_{\tilde{t}_L}$ are 
approximatively the same]. The branching ratios would be, however, 
different because of the different couplings. However, if the mass splitting 
between the two stop eigenstates is sizeable, the additional mode $\tilde{t}_2 
\to \tilde{t}_1 f \bar{f}$ through $Z$ and neutral Higgs boson exchanges has 
to be taken into account. \s

This is shown in Fig.~3 where the decays of $\tilde{t}_2$ are displayed  as a
function of $\tb$, for $\mu=-350$ GeV and $M_2=310$ GeV. We have taken $A_t=A_b
=A_\tau=-100$ GeV and a common slepton mass $m_{\tilde l}=200$ GeV; in the
squark sector, we have used a common mass $m_{\tilde q}=400$ GeV for the first
and second generation squarks  but non--universal masses $m_{\tilde{t}_L}=
m_{\tilde{b}_L}=200$ GeV and $m_{\tilde{t}_R}=120$ GeV to allow for lighter top
squarks with masses $m_{\tilde{t}_1} \sim 200$ and $m_{\tilde{t}_2} \sim 250$
GeV for $\tb \sim 10$ [in this case, $m_{\tilde{t}_2}$ is lighter than the
chargino $\chi_1^+$ so that the two--body decay $\tilde{t}_2 \to b\chi_1^+$ is
shut]. \s 

In this scenario, $\tilde{b}_1$ and the sleptons $\tilde{\tau}_1$ and
$\tilde{\nu}$ are lighter than $\tilde{t}_2$ so that the three--body decays
$\tilde{t}_2 \to \tilde{b}_1 f \bar{f}$ and $\tilde{t}_2 \to b \tilde{\tau}_1
\nu_\tau, b \tilde{\nu}_l l$ [in the figure we sum the branching ratios for all
sleptons] are kinematically  open; the former decay channel is dominant up to
values $\tb \sim 20$ where the two--body decay  $\tilde{t}_2 \to \tilde{b}_1 W$
opens up and reaches a branching fraction close to unity.  The decays into $bW
\chi_1^0$ and $b H^+ \chi_1^0$, as well as the loop induced decay $\tilde{t}_2
\to c \chi_1^0$, are suppressed below the percent level [we cut the branching
ratio for the decay $\tilde{t}_2 \to bW \chi_1^0$, which is mediated by sbottom
exchange, when $\tilde{b}_1$ becomes on--shell since then, we have the decay
chain $\tilde{t}_2 \to \tilde{b}_1 W \to \tilde{b}_1 f\bar{f}$]. The mass
splitting between  $\tilde{t}_2$ and $\tilde{t}_1$ is smaller than the $Z$ and
Higgs boson masses and the three--body decay $\tilde{t}_2 \to \tilde{t}_1
f\bar{f}$  occurs at a sizeable rate for small and intermediate values of
$\tb$, reaching a branching ratio of the order of 50\% for $\tb \sim 15$. \s

\begin{figure}[htbp]
\vspace*{-3.cm}
\hspace*{-3.2cm}
\psfig{figure=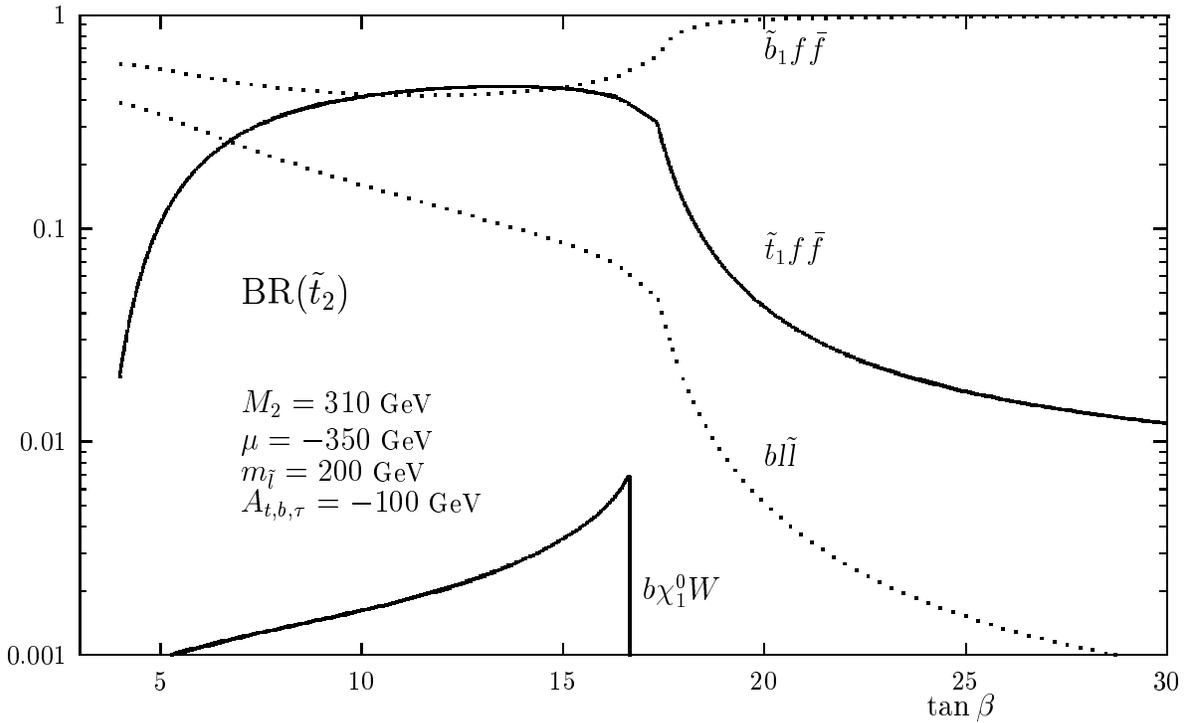,width=21.cm}
\vspace*{-17.5cm}
\caption[]{Example of branching ratios for the two--body and three--body decay 
modes of the heaviest top squark $\tilde{t}_2$ as a function of $\tan \beta$ 
for $\mu=-350$ GeV, $M_2=310$ GeV, $m_{\tilde{l}}=200$ GeV and trilinear
couplings $A_t=A_b=A_\tau=-100$ GeV. \\} 
\end{figure}

Finally, for completeness, we have also studied the three-body decays of
bottoms squarks. In this case, since $\chi_1^0$ is the LSP and $m_b$ is small,
the two--body decay channels $\tilde{b}_i \to b \chi_1^0$ are always
kinematically open so that three--body decays can be hardly competitive. The
only situation where the latter can be sizeable is when the $\tilde{b}_i b
\chi_1^0$ couplings are very tiny. This occurs when the lightest neutralinos
are higgsino--like [one has then to consider both $\chi_1^0$ and $\chi_2^0$
states since in this case, $m_{\chi_1^0} \sim m_{\chi_2^0} \sim |\mu|$] so that
the coupling is suppressed by $m_b/M_W$. One also needs rather small $\tb$
values not to enhance the couplings which are proportional to $1/\cos \beta$.
However, even in this case, the three--body decay $\tilde{b}_1 \to \tilde{t}_1
W^* \to \tilde{t}_1 f\bar{f}'$ for instance has a small branching ratio: it is
only in a rather limited range of the MSSM parameter space that it reaches the
level of a few percent. This is shown in Fig.~4, where for the chosen set of
soft--SUSY breaking parameters, BR($\tilde{b}_1 \to \tilde{t}_1 f\bar{f}'$)
exceeds the percent level only for relatively small values of $\tb$. \s

\begin{figure}[htbp]
\vspace*{-2.5cm}
\hspace*{-2.2cm}
\psfig{figure=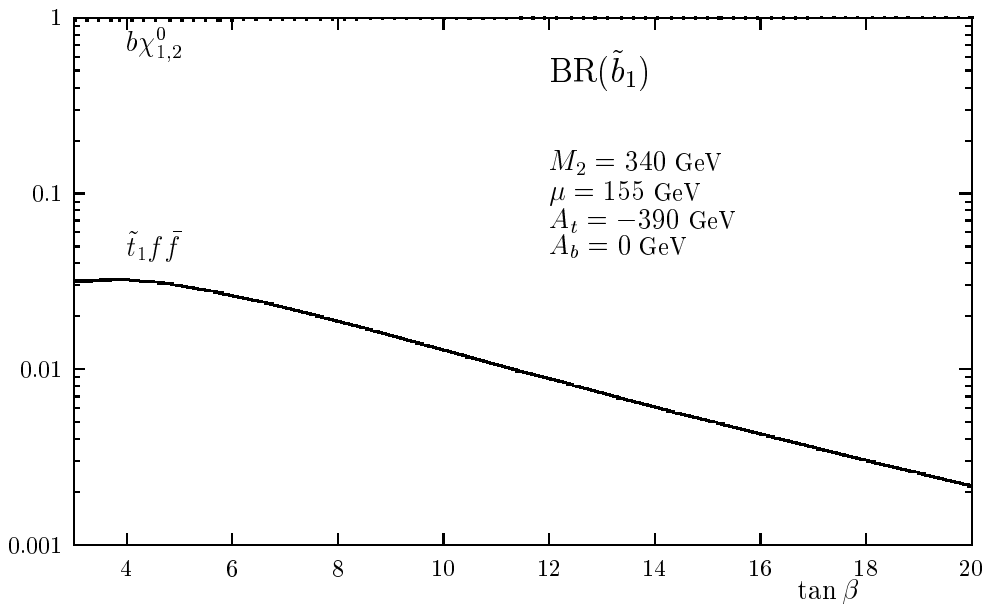,width=19.cm}
\vspace*{-17.3cm}
\caption[]{Example of branching ratios for the two--body and three--body decay 
modes of the lightest bottom squark $\tilde{b}_1$ as a function of $\tan 
\beta$. The soft--SUSY breaking scalar masses are taken to be, $m_{\tilde{t}_L} 
= m_{\tilde{b}_L}=190$ GeV, $m_{\tilde{t}_R}=350$ GeV and 
$m_{\tilde{b}_R}=300$ GeV.}
\end{figure}

The smallness of the three--body decay rates is mainly due to the fact that in
the MSSM one has to take into account the experimental constraints on the
squark masses and on $\tb$ from the negative Higgs boson searches at LEP. In an
unconstrained MSSM, for instance without the unification of the gaugino masses
at the GUT scale [see Ref.~\cite{non-universal}, for examples of models],
several constraints on the SUSY parameters can be relaxed and some three--body
decays might become important. In particular, as it has been recently discussed
in Ref.~\cite{R11}, the decay $\tilde{b}_2 \to \tilde{b}_1 b \bar{b}$ through
virtual gluino exchange can be competitive [since it is a strong interaction
process] with the decay channel $\tilde{b}_2 \to b \chi_1^0$ if the gluino mass
is not too large compared to the lightest neutralino masses, as it might be the
case in the models discussed in Ref.~\cite{non-universal}.

\subsection*{6. Conclusions}

We have performed a comprehensive analysis of the decays of third generation
squarks, focusing on the three body decay modes. Because of the large value of
the top quark mass and the possible large mixing in the stop and sbottom
sectors which leads to sizeable splitting between the masses of the two
physical states, the decay pattern of scalar top [and to a lesser extent  
bottom] quarks can be dramatically different from the decay pattern of first 
and second generation squarks, which simply decay into their almost massless 
partner quarks and neutralino  or chargino states. \s

Several new decay channels, including the cascade decays of heavier squarks
into lighter ones and fermion--antifermion pairs as well the decays of both top
squarks into $W, H^+$ bosons and the lightest neutralinos or into leptons and
lighter sleptons, are possible.  In some areas of the MSSM parameter space,
these additional decay modes can have sizeable branching fractions, and they
can even be the dominating decay modes. These channels need therefore to be
taken into account in the search of scalar top and bottom quarks at present and
future colliders. 

\bigskip

\nn {\bf Acknowledgments:} \s

\nn We thank the members of the French GDR--Supersym\'etrie, in particular 
Jean-Fran\c cois Grivaz, as well as Aseshkrishna Datta for various discussions. 
Discussions with Werner Porod on Refs.~\cite{R6a,R6b} are gratefully 
acknowledged. This work is partially supported by the European Union under 
contract HPRN-CT-2000-00149.


\begin{thebibliography}{99}
\bibitem{sdecay} Y. Mambrini, {\tt SDECAY}: a program for the decays of SUSY 
particles, to be released.
\end{thebibliography}
\end{document}